\documentclass[preprint,12pt]{revtex4}
\usepackage{graphicx}% Include figure files

\begin{document}

\title[Diagnosis and Prediction of Market Rebounds in Financial Markets]{Diagnosis and Prediction of Market Rebounds in Financial Markets}

\author{Wanfeng Yan $\dag$, Ryan Woodard $\dag$ and Didier Sornette $\dag \ddag$}
\email{wyan@ehtz.ch, rwoodard@ethz.ch, dsornette@ethz.ch}
\thanks{\\~Corresponding author: Didier Sornette.}
\address{\normalfont{$\dag$ Chair of Entrepreneurial Risks\\
Department of Management, Technology and Economics\\ ETH Z\"{u}rich,
CH-8001 Z\"{u}rich, Switzerland \vspace{12pt}\\ $\ddag$ Swiss
Finance Institute, c/o University of Geneva\\ 40 blvd. Du Pont
dArve, CH 1211 Geneva 4, Switzerland}\\\vspace{12pt}}

\begin{abstract}
We introduce the concept of ``negative bubbles'' as the mirror (but
not necessarily exactly symmetric) image of standard financial
bubbles, in which positive feedback mechanisms may lead to transient
accelerating price falls. To model these negative bubbles, we adapt
the Johansen-Ledoit-Sornette (JLS) model of rational expectation
bubbles with a hazard rate describing the collective buying pressure
of noise traders. The price fall occurring during a transient
negative bubble can be interpreted as an effective random down
payment that rational agents accept to pay in the hope of profiting
from the expected occurrence of a possible rally. We validate the
model by showing that it has significant predictive power in
identifying the times of major market rebounds. This result is
obtained by using a general pattern recognition method that combines
the information obtained at multiple times from a dynamical
calibration of the JLS model. Error diagrams, Bayesian inference and
trading strategies suggest that one can extract genuine information
and obtain real skill from the calibration of negative bubbles with
the JLS model. We conclude that negative bubbles are in general
predictably associated with large rebounds or rallies, which are the
mirror images of the crashes terminating standard bubbles.

\end{abstract}

\maketitle \clearpage

\section{Introduction}

Financial bubbles are generally defined as transient upward acceleration of prices
above fundamental value \cite{Galbraith,Kindleberger,sornettecrash}.
However, identifying unambiguously the presence of a bubble remains
an unsolved problem in standard econometric and  financial
economic approaches  \cite{gurkaynak-2008,Lux-Sornette},
due to the fact that the fundamental value is in general poorly constrained
and it is not possible to distinguish between
exponentially growing fundamental price and exponentially growing bubble price.

To break this stalemate, Sornette and co-workers have proposed
that bubbles are actually not characterized by exponential prices (sometimes
referred to as ``explosive''),
but rather by faster-than-exponential growth of price (that should therefore
be referred to as ``super-explosive''). See \cite{sornettecrash}
and references therein. The reason for such faster-than-exponential regimes
is that imitation and herding behavior of noise traders
and of boundedly rational agents create positive feedback in
the valuation of assets, resulting in price processes that exhibit
a finite-time singularity at some future time $t_c$. See \cite{GorielyHyde00}
for a general theory of finite-time singularities in ordinary differential equations,
\cite{GluzmanSornette02} for a classification
and \cite{JohansenSornette01,SammisSornette02,SorTakaZhou03} for applications. This critical time $t_c$
is interpreted as the end of the bubble, which is often but not necessarily the time
when a crash occurs \cite{Johansen-Sornette}. Thus, the main difference
with standard bubble models is that the underlying price process is
considered to be intrinsically transient due to positive feedback mechanisms
that create an unsustainable regime. Furthermore, the tension and competition between the
value investors and the noise traders may create deviations around the finite-time singular
growth in the form of oscillations that are periodic in the logarithm of the time to $t_c$.
Log-periodic oscillations appear to our clocks as peaks and valleys with
progressively greater frequencies that eventually reach
a point of no return, where the unsustainable growth has the highest
probability of ending in a violent crash or gentle deflation of the bubble.
Log-periodic oscillations are associated with the symmetry of discrete scale
invariance, a partial breaking of the symmetry of continuous scale invariance,
and occurs in complex systems characterized by a hierarchy of scales. See
\cite{DSI-sornette98}  for a general review and references therein.

Recent literatures on bubbles and crashes can be summarized as the
following kinds: first, the combined effects of heterogeneous
beliefs and short-sales constraints may cause large movements in
asset. In this kind of models, the asset prices are determined at
equilibrium to the extent that they reflect the heterogeneous
beliefs about payoffs. But short sales restrictions force the
pessimistic investors out of the market, leaving only optimistic
investors and thus inflated asset price levels. However, when short
sales restrictions no longer bind investors, then prices fall back
down
\cite{lintner,miller,harrisonkreps,jarrow,chen,scheinkmanxiong,duffie,abreubrunnermeier}.
While in the second type, the role of ``noise traders'' in fostering
positive feedback trading has been emphasized. These models says
trend chasing by one class of agents produces momentum in stock
prices \cite{delong,barberis,daniel,hong}. The empirical evidence on
momentum strategies can be found in
\cite{jegadeeshtitman,jegadeeshtitman01,debondtthaler}.

After the discussion on bubbles and crashes, the literatures on
rebound should be summarized also. On the theoretical side, there
are several competing explanations for price decreases followed by
reversals: liquidity and time-varying risk. \cite{pedersen} stresses
the importance of liquidity: as more people sell, agents who
borrowed money to buy assets are forced to sell too. When forced
selling stops, this trend reverses. \cite{nagel} shows that it is
risky to be a fundamental trader in this environment and that price
reversals after declines are likely to be higher when there is more
risk in the price, as measured by volatility. On the empirical
front concerning the forecast of reversals in price drops, \cite{jegadeeshtitman}
shows that the simplest way to predict prices is to look at past
performance.  \cite{campbellshiller88} shows that price-dividend
ratios forecast future returns for the market as a whole. However,
these two approaches do not aim at predicting and cannot
determine the most probable rebound time
for a single ticker of the stock. The innovation of our methodology in this
respect is to provide a very detailed method to detect rebound
of any given ticker.

In this paper, we explore the hypothesis that financial bubbles have
mirror images in the form of ``negative bubbles'' in which positive
feedback mechanisms may lead to transient accelerating price falls.
We adapt the Johansen-Ledoit-Sornette (JLS)  model of rational
expectation bubbles \cite{js,jsl,jls}  to negative bubbles. The
crash hazard rate becomes the rally hazard rate, which quantifies
the probability per unit time that the market rebounds in a strong
rally. The upward accelerating bullish price characterizing a
bubble, which was the return that rational investors require as a
remuneration for being exposed to crash risk, becomes a downward
accelerating bearish price of the negative bubble, which can be
interpreted as the cost that rational agents accept to pay to profit
from a possible future rally. During this accelerating downward
trend, a tiny reversal could be a strong signal for all the
investors who are seeking the profit from the possible future rally.
These investors will long the stock immediately after this tiny
reversal. As a consequence, the price rebounds very rapidly.

This paper contributes to the literature by augmenting the evidence
for transient pockets of predictability that are characterized by
faster-than-exponential growth or decay. This is done by adding the
phenomenology and modeling of ``negative bubbles'' to the evidence
for characteristic signatures of (positive) bubbles.  Both positive
and negative bubbles are suggested to result from the same
fundamental mechanisms, involving imitation and herding behavior
which create positive feedbacks. By such a generalization within the
same theoretical framework, we hope to contribute to the development
of a genuine science of bubbles.

The rest of the paper is organized as follows. Section 2.1
summarizes the main definitions and properties of the
Johansen-Ledoit-Sornette (JLS) for (positive) bubbles and their
associated crashes. Section 2.2 presents the modified JLS model for
negative bubbles and their associated rebounds (or rallies). The
subsequent sections test the JLS model for negative bubbles by
providing different validation steps, in terms of prediction skills
of actual rebounds and of abnormal returns of trading strategies
derived from the model. Section 3 describes the method we have
developed to test whether the adapted JLS model for negative bubbles
has indeed skills in forecasting large rebounds. This method uses a
robust pattern recognition framework build on the information
obtained from the calibration of the adapted JLS model to the
financial prices. Section 4 presents the results of the tests
concerning the performance of the method of section 3 with respect
to the advanced diagnostic of large rebounds. Section 5 develops
simple trading strategies based on the method of section 3, which
are shown to exhibit statistically significant returns, when
compared with random strategies without skills with otherwise
comparable attributes. Section 6 concludes.

\section{Theoretical model for detecting rebounds}

\subsection{Introduction to the JLS model and bubble conditions}

\cite{js}, \cite{jsl}, \cite{jls} developed a model (referred to below as the JLS model) of
financial bubbles and crashes, which is an extension of the rational expectation
bubble model of \cite{Blanchardwat}. In this model,
a crash is seen as an event potentially
terminating the run-up of a bubble. A financial bubble is modeled as a regime of
accelerating (super-exponential power law) growth punctuated by short-lived
corrections organized according the symmetry of
discrete scale invariance \cite{DSI-sornette98}. The super-exponential power law is argued
to result from positive feedback resulting from noise trader decisions that tend
to enhance deviations from fundamental valuation in an accelerating spiral.

In the JLS model, the dynamics of stock
markets is described as
\begin{equation}
  \frac{dp}{p} = \mu(t)dt + \sigma(t)dW - \kappa dj~,
  \label{eq:dynamic}
\end{equation}
where $p$ is the stock market price, $\mu$ is the drift (or trend) and $dW$ is
the increment of a Wiener process (with zero mean and unit variance). The term $dj$
represents a discontinuous
jump such that $dj = 0$ before the crash and $dj = 1$ after the crash occurs.
The loss amplitude associated with the occurrence of a crash is determined by the parameter $\kappa$.
The assumption of the constant jump size is easily relaxed by considering
a distribution of jump sizes, with the condition that its first moment exists. Then, the no-arbitrage condition
is expressed similarly with $\kappa$ replaced by its mean.
Each successive crash corresponds to a jump of $dj$ by one unit. The
dynamics of the jumps is governed by a crash hazard rate $h(t)$.
Since $h(t) dt$ is the probability that the crash occurs between $t$ and $t+dt$ conditional
on the fact that it has not yet happened,
we have $ E_t[dj]  = 1 \times h(t) dt + 0 \times (1- h(t) dt)$ and therefore
\begin{equation}
  {\rm E}_t[dj] = h(t)dt~.
  \label{theyjytuj}
\end{equation}
Under the assumption of the JLS model, noise traders exhibit collective
herding behaviors that may destabilize the market. The JLS model
assumes that the aggregate effect of noise traders can be accounted
for by the following dynamics of the crash hazard rate
\begin{equation}
  h(t) = B'(t_c-t)^{m-1}+C'(t_c-t)^{m-1}\cos(\omega\ln (t_c-t) -\phi')~.
  \label{eq:hazard}
\end{equation}
The intuition behind this specification (\ref{eq:hazard}) has been presented at length in
\cite{js,jsl,jls}, among others, and further developed in (Sornette and Johansen, 2002)
for the power law part and by  \cite{IdeSornette} and (Zhou et al., 2005) for the
second term in the right-hand-side of expression (\ref{eq:hazard}).
In a nutshell, the power law behavior $\sim t_c-t)^{m-1}$ embodies the
mechanism of positive feedback posited to be at the source of the bubbles.
If the exponent $m<1$, the crash hazard may diverge as $t$ approaches
a critical time $t_c$, corresponding to the end of the bubble.
The cosine term in the r.h.s. of (\ref{eq:hazard}) takes into account
the existence of a possible hierarchical cascade of  panic acceleration
punctuating the course of the bubble, resulting either from a preexisting
hierarchy in noise trader sizes \cite{SornetteJohansen97} and/or from the interplay between market price
impact inertia and nonlinear fundamental value investing  \cite{IdeSornette}.

The no-arbitrage condition reads ${\rm E}_t[dp]=0$, where the expectation is performed
with respect to the risk-neutral measure, and in the frame of the risk-free rate. This is the
standard condition that the price process is a martingale. Taking the expectation of expression (\ref{eq:dynamic})
under the filtration (or history) until time $t$ reads
\begin{equation}
{\rm E}_t[dp]= \mu(t) p(t) dt + \sigma(t) p(t) {\rm E}_t[dW] - \kappa p(t) {\rm E}_t[dj]~.
 \label{thetyjye}
\end{equation}
Since ${\rm E}_t[dW] =0$ and  ${\rm E}_t[dj] = h(t)dt$ (equation (\ref{theyjytuj})),
together with the no-arbitrage condition ${\rm E}_t[dp]=0$,
this yields
\begin{equation}
\mu(t) = \kappa h(t)~.
\label{tjyj4n}
\end{equation}
This result (\ref{tjyj4n}) expresses that the return $\mu(t)$ is controlled by the risk of the crash
quantified by its crash hazard rate $h(t)$.

Now, conditioned on the fact that no crash occurs,
equation (\ref{eq:dynamic}) is simply
\begin{equation}
  \frac{dp}{p} = \mu(t)dt + \sigma(t)dW = \kappa h(t) dt + \sigma(t)dW~.
  \label{eq:dynaiuhomic}
\end{equation}
Its conditional expectation leads to
\begin{equation}
{\rm E}_t\left[\frac{dp}{p}\right] =  \kappa h(t) dt~.
\end{equation}
Substituting with the expression (\ref{eq:hazard}) for $h(t)$ and integrating
yields the so-called log-periodic power law (LPPL) equation:
\begin{equation}
  \ln {\rm E}[p(t)] = A + B(t_c-t)^m + C(t_c-t)^m\cos(\omega\ln (t_c-t) - \phi)
\label{eq:lppl}
\end{equation}
where $ B = - \kappa B' /m$ and $C = - \kappa C' /
\sqrt{m^2+\omega^2}$. Note that this expression (\ref{eq:lppl})
describes the average price dynamics only up to the end of the
bubble. The JLS model does not specify what happens beyond $t_c$.
This critical $t_c$ is the termination of the bubble regime and the
transition time to another regime. This regime could be a big crash
or a change of the growth rate of the market. Merrill Lynch EMU
(European Monetary Union) Corporates Non-Financial Index in 2009
\cite{BFE-FCO09} provides a vivid example of a change of regime
characterized by a change of growth rate rather than by a crash or rebound.
 For $m<1$, the crash hazard rate accelerates up to $t_c$ but
its integral up to $t$ which controls the total probability for a
crash to occur up to $t$ remains finite and less than $1$ for all
times $t \leq t_c$. It is this property that makes it rational for
investors to remain invested knowing that a bubble is developing and
that a crash is looming. Indeed, there is still a finite probability
that no crash will occur during the lifetime of the bubble. The
excess return $\mu(t) = \kappa h(t)$ is the remuneration that
investors require to remain invested in the bubbly asset,  which is
exposed to a crash risk. The condition that the price remains finite
at all time, including $t_c$, imposes that $m > 0$.

Within the JLS framework, a bubble is qualified when
the crash hazard rate accelerates. According to (\ref{eq:hazard}), this
imposes $m<1$ and $B'>0$, hence $B<0$ since $m > 0$ by the condition
that the price remains finite. We thus have a first condition for a bubble to occur
\begin{equation}
  0 < m < 1~.
\label{eq:m}
\end{equation}
By definition, the crash rate should be non-negative. This imposes \cite{bm}
\begin{equation}
 b \equiv -Bm - |C|\sqrt{m^2+\omega^2}  \geq 0~.
  \label{eq:bg0}
\end{equation}

\subsection{Modified JLS model for ``negative bubbles'' and rebounds}

As recalled above, in the JLS framework,
financial bubbles are defined as transient regimes of faster-than-exponential
price growth resulting from positive feedbacks. We refer to these
regimes as ``positive bubbles.''
We propose that positive
feedbacks leading to increasing amplitude of the price momentum can also occur in
a downward price regime and that
transient regimes of faster-than-exponential {\it downward} acceleration can exist.
We refer to these regimes as ``negative bubbles.''
In a ``positive'' bubble regime, the larger the price is, the larger the increase
of future price. In a ``negative bubble'' regime, the smaller the price, the
larger is the decrease of future price. In a positive bubble, the positive feedback
results from over-optimistic expectations of future returns leading to self-fulfilling
but transient unsustainable price appreciations. In a negative bubble,
the positive feedbacks reflect the rampant pessimism fueled by short positions
leading investors to run away from the market which spirals downwards also
in a self-fulfilling process.

The symmetry between positive and negative bubbles is obvious for
currencies. If a currency A appreciates abnormally against another currency B
following a faster-than-exponential trajectory, the value of currency B expressed
in currency A will correspondingly fall faster-than-exponentially in a downward spiral.
In this example, the negative bubble is simply obtained by taking the
inverse of the price, since the value of currency A in units of B is the inverse
of the value of currency B in units of A. Using logarithm of prices, this corresponds
to a change of sign, hence the ``mirror'' effect mentioned above.

The JLS model provides a suitable framework to describe negative bubbles,
with the only modifications
that both the expected excess return $\mu(t)$ and the crash amplitude
$\kappa$ become negative (hence the term ``negative'' bubble). Thus, $\mu$ becomes the expected (negative)
return (i.e., loss) that investors accept to bear, given that they anticipate a potential
rebound or rally of amplitude $|\kappa|$. Symmetrically to the case of positive bubbles,
the price loss before the potential rebound plays the role of a random payment that the
investors honor in order to remain invested and profit from the possible rally.
The hazard rate $h(t)$ now
describes the probability per unit time for the rebound to occur.
The fundamental equations (\ref{eq:hazard}) and (\ref{eq:lppl}) then hold
mutatis mutandis with the inequalities
\begin{equation}
  B > 0 ~,~~~~b < 0 \label{eq:rb}
\end{equation}
being the opposite to those corresponding to a positive bubble
as described in the preceding subsection.

An example of the calibration of a negative bubble with the
JLS model (4) to the S\&P 500 index from 1973-01-01 to 1974-10-01 is
shown in the upper panel of Figure 1. During this period, the S\&P
500 index decreased at an accelerating pace. This price fall was
accompanied by very clear oscillations that are log-periodic in
time, as described by the cosine term in formula (4). Notice that
the end of the decreasing market is followed by a dramatic rebound
in index price. We hypothesize that, similar to a crash following an
unsustainable super-exponential price appreciation (a positive
bubble), an accelerating downward price trajectory (a negative
bubble) is in general followed by a strong rebound. Furthermore, in
order to suggest that this phenomenon is not an isolated phenomenon
but actually  happens widely in all kinds of
markets, another example in the foreign exchange market is presented in the
lower panel of Figure 1. The USD/EUR change rate from 2006-07-01 to
2008-04-01 also underwent a significant drawdown with very clear
log-periodic oscillations, followed by a strong positive rebound.
One of the goals of this paper is to identify such regions of
negative bubbles in financial time series and then use a pattern
recognition method to distinguish ones that were (in a back-testing
framework) followed by significant price rises.

In financial markets, large
positive returns are less frequent than large negative returns, as expressed
for instance in the skewness of return distributions. However, when
studying drawdowns and drawups (i.e., runs of same sign returns).
Johansen and Sornette found that, for
individual companies, there are approximately twice as many large
rallies as crashes with amplitude larger than 20\% with durations of a few
days \cite{jsdrawdown}.

\section{Rebound prediction method}

We adapt the pattern recognition method of \cite{gel}
to generate predictions of rebound times in financial markets on the basis
of the detection and calibration of negative bubbles, defined in the previous section.
We analyze the
S\&P 500 index prices, obtained from Yahoo! finance for ticker `\^{ }GSPC'
(adjusted close price)\footnote{http://finance.yahoo.com/q/hp?s=\^{ }GSPC}.
The start time of our time series is 1950-01-05, which is very close to the
first day when the S\&P 500 index became available (1950-01-03). The last day of our tested time
series is 2009-06-03.

\subsection{Fitting methods}

We first divide our S\&P 500 index time series into different sub-windows
$(t_1, t_2)$ of length $dt \equiv t_2 - t_1$ according to the following rules:
\begin{enumerate}
\item The earliest start time of the windows is $t_1 = $ 1950-01-03.  Other
  start times $t_1$ are calculated using a step size of $dt_1 = 50$ calendar
  days.
\item The latest end time of the windows is $t_2 = $ 2009-06-03.  Other end
  times $t_2$ are calculated with a negative step size $dt_2 = -50$ calendar
  days.
\item The minimum window size $dt_{\mathrm{min}} = 110$ calendar days.
\item The maximum window size $dt_{\mathrm{max}} = 1500$ calendar days.
\end{enumerate}
These rules lead to 11,662 windows in the S\&P 500 time series.

For each window, the log of the S\&P 500 index is fit with the
JLS equation (\ref{eq:lppl}). The fit is performed in two steps.
First, the linear parameters $A, B$ and $C$ are
slaved to the non-linear parameters by solving them analytically as a function
of the nonlinear parameters. We refer to \cite{jls} (page 238 and following ones), which gives
the detailed equations and procedure.
Then, the search space is obtained as a 4 dimensional
parameter space representing $m, \omega, \phi, t_c$.  A heuristic search
implementing the Tabu algorithm \cite{ck} is used to find initial
estimates of the parameters which are then passed to a Levenberg-Marquardt
algorithm \cite{kl,dm} to minimize the residuals (the sum
of the squares of the differences) between the model and the data.  The bounds
of the search space are:
\begin{eqnarray}
  m &\in& [0.001, 0.999] \label{eq:rangem}\\
  \omega &\in& [0.01, 40] \\
  \phi &\in& [0.001, 2\pi]\\\label{eq:rangetc}
  t_c &\in& [t_2, t_2 + 0.375(t_2 - t_1)]
\end{eqnarray}
We choose these bounds because $m$ has to be between $0$ and $1$
according to the discussion before; the log-angular frequency
$\omega$ should be greater than $0$. The upper bound $40$ is large
enough to catch high-frequency oscillations (though we later discard
fits with $\omega > 20$); phase $\phi$ should be between 0 and
$2\pi$; as we are predicting a critical time in financial markets,
the critical time should be after the end of the time series we are
fitting. Finally, the upper bound of the critical time should not be
too far away from the end of the time series since predictive
capacity degrades far beyond $t_2$. We have empirically found
elsewhere \cite{Jiangetal09} one-third of the interval width to be a
good cut-off.

The combination of the heuristic and optimization results in a set
of parameters $A, B, C, m, \omega, \phi$ and $t_c$ for each of the
11,662 windows.  Of these parameter sets, 2,568 satisfy the negative
bubble condition (\ref{eq:rb}). In Figure~\ref{fg:histo}, we plot
the histogram of critical time $t_c$ for these negative bubble fits
and the \emph{negative} logarithm of the S\&P 500 time series. Peaks
in this time series, then, indicate minima of the prices, many of
these peaks being preceded by a fast acceleration with upward
curvature indicating visually a faster-than-exponential growth of $-
p(t)$. This translates into accelerating downward prices. Notice
that many of these peaks of $-\ln p(t)$ are followed by sharp drops,
that is, fast rebounds in the regular $+\ln p(t)$. We see that peaks
in $-\ln p(t)$ correspond to peaks in the negative bubble $t_c$
histogram, implying that the negative bubbles qualified by the JLS
model are often followed by rebounds.  This suggests the possibility
to diagnose negative bubbles and their demise in the form of a
rebound or rally. If correct, this hypothesis would extend the
proposition \cite{Jiangetal09,BFE-FCO10}, that financial bubbles can
be diagnosed before their end and their termination time can be
determined with an accuracy better than chance, to negative bubble
regimes associated with downward price regimes. We quantify this
observation below.

\subsection{Definition of rebound}
\label{defreboundh2ysec}

The aim is first to recognize different patterns in the S\&P 500
index from the 11,662 fits and then use the subset of 2,568 negative
bubble fits to identify specific negative bubble characteristics.
These characteristics will then be used to `predict' (in a
back-testing sense) negative bubbles and rebounds in the future.

We first define a rebound, note as $ \mathrm{Rbd}$. A day $d$ is a rebound $ \mathrm{Rbd}$ if the price on that day is the
minimum price in a window of 200 days before and 200 days after it.  That is,
\begin{equation}
  \mathrm{Rbd} = \{d \mid P_d = \min\{P_x\}, \forall x \in [d-200,d+200]\}~
  \label{defreboundh2y}
\end{equation}
where $P_d$ is the adjusted closing price on day $d$.
We find 19 rebounds of the $\pm 200$-days type\footnote{Ten rebounds
in the back tests before 1975.1.1: 1953-09-14;
1957-10-22;   1960-10-25;  1962-06-26; 1965-06-28; 1966-10-07;
 1968-03-05;  1970-05-26;  1971-11-23; 1974-10-03 and nine rebounds after 1975.1.1
 in the prediction range: 1978-03-06; 1980-03-27; 1982-08-12; 1984-07-24; 1987-12-04;
 1990-10-11; 1994-04-04; 2002-10-09; 2009-03-09.}
in the 59 year S\&P 500 index history. Our task is to diagnose such
rebounds in advance. We could also use other numbers instead of 200
to define a rebound. The predictability is stable with respect to a
change of this number. This is because we learn from the learning
set with a certain number type of rebounds and try to predict the
rebounds of the same type. Later we will also show the results for
$\pm 365$-days type of rebounds.

\subsection{Definitions and concepts needed to set up the pattern recognition method}
\label{sec:prelude}

In what follows we describe a hierarchy of descriptive and quantitative terms as follows.
\begin{itemize}
\item \textbf{learning set.} A subset of the whole set which only contains the fits with critical times in the past. We learn the properties of historical rebounds from this set and develop the predictions based on these properties.
\item \textbf{classes.} Two classes of fits are defined according to whether
the critical time of a given fit is near some rebound or not, where `near' will be defined below.
\item \textbf{groups.} A given group contains all fits of a given window size.
\item \textbf{informative parameters.} Informative parameters are the distinguishing parameters of fits in the same group but different classes.
\item \textbf{questionnaires.} Based on the value of an informative parameter, one can ask if a certain trading day is a start of rebound or not. The answer series generated by all the informative parameters is called questionnaire.
\item \textbf{traits.} Traits are extracted from questionnaire. They are short and contain crucial information and properties of a questionnaire.
\item \textbf{features.} Traits showing the specific property of a single class are selected to be the feature of that class.
\item \textbf{rebound alarm index.} An index developed from features to show the probability that a certain day is a rebound.
\end{itemize}

In this paper, we will show how all the above objects are constructed. Our final goal is to make predictions for the rebound time. The development of the rebound alarm index will enable us to achieve our goal. Several methodologies are presented to quantify the performance of the predictions.

\subsection{Classes}

In the pattern recognition method of \cite{gel}, one should define the learning set to find
characteristics that will then be used to make predictions. We designate all
fits before Jan. 1, 1975 as the learning set $\Sigma_1$:
\begin{equation}
  \Sigma_1 = \{f \mid t_{c,f}, t_{2,f} < Jan. 1, 1975\}
\end{equation}
There are 4,591 fits in this set, which we all use without any pre-selection.
No pre-selection for instance
using Eq.~(\ref{eq:rb})  is applied, on the basis of  the robustness of the pattern recognition method.
We then distinguish two different classes from $\Sigma_1$ based on the critical
time $t_c$ of the fits. For a single fit $f$ with critical time $t_{c,f}$, if
this critical time is within $D$ days of a rebound, then we assign fit $f$ to
Class I, represented by the symbol $C_I$. Otherwise, $f$ is assigned to Class II,
represented by the symbol $C_{II}$.  For this
study, we chose $D = 10$ days because $D$ too big will lose precision and $D$ too small will take the noise into account.
In this case, Class I fits are those with $t_c$
within 10 days of one of the 19 rebounds. We formalize this rule as:
\begin{eqnarray}
  C_I & = & \{f \mid f \in \Sigma_1, \exists d \in \mathrm{Rbd}, s.t. |t_{c,f} - d|
    \leq D \}, \\
  C_{II} & = & \{f \mid f \in \Sigma_1, |t_{c,f} - d| > 10, \forall d \in
    \mathrm{Rbd}\}, \\
  D & = & 10~ \mathrm{days}.
\end{eqnarray}
To be clear, Class I is formed by all the fits in learning set $\Sigma_1$ which
have a critical time $t_c$ within 10 days of one of the rebounds.  All of the
fits in the learning set which are not in Class I are in Class II.

\subsection{Groups}
\label{sec:groups}

We also categorize all fits into separate \emph{groups} (in addition to the two
\emph{classes} defined above) based on the length of the fit interval, $L_f =
dt = t_2 - t_1$.  We generate 14 groups, where a given group $G_i$ is defined
by:
\begin{equation}
  G_i = \{f \mid L_f \in [100i, 100i + 100], i = 1, 2, ..., 14, f \in
  \Sigma_1\}
  \label{eq:groups}
\end{equation}
All 4,591 fits in the learning set are placed into one of these 14 groups.

\subsection{Informative Parameters}
\label{sec:ip}

For each fit in the learning set, we take 6 parameters to construct a flag that determines
the characteristics of classes. These 6 parameters are $m, \omega, \phi$ and
$B$ from Eq.~(\ref{eq:lppl}), $b$ (the negative bubble condition) from
Eq.~(\ref{eq:bg0}) and $q$ as the residual of the fit.

We categorize these sets of 6 parameters for fits which are in the same group
and same class.  Then for each class-group combination, we calculate the
probability density function (pdf) of each parameter using the adaptive kernel
method \cite{bw}, generating 168 pdfs (6 parameters $\times$ 2 classes
$\times$ 14 groups).

We compare the similarity (defined below) of the pdfs of each of the six
parameters that are in the same group (window length) but different classes
(proximity of $t_c$ to a rebound date). If these two pdfs are similar, then we
ignore this parameter in this group. If the pdfs are different, we record this
parameter of this group as an \emph{informative parameter}. The maximum number
of possible informative parameters is 84 (6 parameters $\times$ 14 groups).

We use the Kolmogorov-Smirnov method \cite{ks} to detect the difference
between pdfs. If the maximum difference of the cumulative distribution
functions (integral of pdf) between two classes exceeds 5\%, then this is an
informative parameter.  We want to assign a uniquely determined integer $IP_l$
to each informative parameter.  We can do so by using three indexes, $i, j$ and
$l$.  The index $i$ indicates which group, with $i \in [1 ,14]$.  The index $j$
indicates the parameter, where $j = 1,2,3,4,5,6$ refer to $m, \omega, \phi, B,
b, q,$ respectively.  Finally, $l$ represents the actual informative parameter.
Assuming that there are $L$ informative parameters in total
and using the indexes, $IP_l$ is then calculated via
\begin{equation}
  IP_l = 6i + j
  \label{eq:IP}
\end{equation}
for $l \in [1, L]$.

Given the $L$ informative parameters $IP_l$, we consider the pdfs
for the two different classes of a single informative parameter.
The set of abscissa values within the allowed range given by
equations (\ref{eq:rangem} - \ref{eq:rangetc}), for which the pdf of
Class I is larger than the pdf of Class II, defines the domain
$Rg_{I,l}$ (`good region') of this informative parameter which is
associated with Class I. The other values of the informative
parameters for which the pdf of Class I is smaller than the pdf of
Class II define the domain $Rg_{II,l}$ which is associated with
Class II. These regions play a crucial role in the generation of
questionnaires in the next section.

Our hypothesis is that
many ``positive'' and ``negative bubbles'' share the same
structure described by the JLS model, because they result
from the same underlying herding mechanism. However,
nothing a priori imposes that the control parameters should be
identical. Note that our pattern recognition methodology
specifically extract the typical informative
parameter ranges that characterize the ``negative bubbles''.

\subsection{Intermediate summary}
\label{sec:intermediate-summary}

We realize that many new terms are being introduced, so in an attempt to be
absolutely clear, we briefly summarize the method to this point.  We sub-divide
a time series into many windows $(t_1, t_2)$ of length $L_f = t_2 - t_1$.  For
each window, we obtain a set of parameters that best fit the model
(\ref{eq:lppl}).  Each of these windows will be assigned one of two
\emph{classes} and one of 14 \emph{groups}.  Classes indicate how close the
modeled critical time $t_c$ is to a historical rebound, where Class I indicates
`close' and Class II indicates `not close'.  Groups indicate the length of the
window.  For each fit, we create a set of six parameters: $m, \omega, \phi$ and
$B$ from Eq.~(\ref{eq:lppl}), $b$ (the negative bubble condition) from
Eq.~(\ref{eq:bg0}) and $q$ as the residual of the fit.  We create the
pdfs of each of these parameters for each fit and define \emph{informative
parameters} as those parameters for which the pdfs differ significantly according to a
Kolmogorov-Smirnov test.  For each informative parameter, we find the regions of
the abscissa of the pdf for which the Class I pdf (fits with $t_c$ close to a
rebound) is greater than the Class II pdf.  For informative parameter $l$
(defined in (\ref{eq:IP})), this region is designated as $Rg_{I,l}$.  In the next section, we will
use these \emph{regions} to create \emph{questionnaires} that will be used to
predictively identify negative bubbles that will be followed by rebounds.

Another important distinction to remember at this point is that the above
method has been used to find \emph{informative parameters} that will be used
below.  Informative parameters are associated with a \emph{class} and a
\emph{group}.

\subsection{Questionnaires}\label{sub:question}

Using the informative parameters and their pdfs described above, we can
generate \emph{questionnaires} for each day of the learning or testing set.
Questionnaires will be used to identify negative bubbles that will be followed
by rebounds.  The algorithm for generating questionnaires is the following:

\begin{enumerate}
\item Obtain the maximum ($t_{c\mathrm{max}}$) and minimum
  ($t_{c\mathrm{min}}$) values of $t_c$ from some subset $\Sigma_{sub}$,
  either the `learning' set or the `predicting (testing)' set of all 11,662 fits.
\item Scan each day $t_{scan}$ from $t_{c\mathrm{min}}$ to
  $t_{c\mathrm{max}}$.  There will be $N = t_{c\mathrm{max}} -
  t_{c\mathrm{min}} + 1$ days to scan.  For each scan day, create a new set
  $S_{t_{scan}}$ consisting of \emph{all} fits in subset $\Sigma_{sub}$ that
  have a $t_c$ near the scan day $t_{scan}$, where `near' is defined using the
  same criterion used for defining the two classes, namely $D = 10$ days:
  \begin{equation}
    S_{t_{scan}} = \{ f \mid |t_{c,f} - t_{scan}| \leq D, f \in \Sigma_{sub} \}
    \label{eq:near}
  \end{equation}
  The number $\# S_{t_{scan}}$ of fits in each set  can be $0$ or greater.
  The sum of the number of fits found in all of the sets
  $\sum_{t_{scan} = t_{c\mathrm{min}}}^{t_{c\mathrm{max}}} \# S_{t_scan}$
  can actually be greater than the total number of fits in $\Sigma_{sub}$ since some fits
  can be in multiple sets.
  Notice that the fits in each set $S_{t_{scan}}$ can (and do) have varying
  window lengths.  At this point, only the proximity to a scan day is used
  to determine inclusion in a scan set.
\item Assign a group to each of the fits in $S_{t{scan}}$.  Recall that groups
  are defined in Eq.~(\ref{eq:groups}) and are based on the window length $L_f
  = dt = t_2 - t_1$.
\item Using all sets $S_{t{scan}}$, for each informative parameter $IP_l$ found
  in Sec.~\ref{sec:ip}, determine if it belongs to Class I (close to a rebound)
  or Class II (not close to a rebound). There are 3 possible answers: 1 =
  `belongs to Class I', -1 = `belongs to Class II' or 0 = `undetermined'.
\end{enumerate}

The status of `belonging to Class I' or not is determined as follows.  First,
find all values of the informative parameter $IP_l$ in a particular scan set
$S_{t_{scan}}$.  For instance, if for a particular scan day $t_{scan}$, there
are $n$ fits in the subset $\Sigma_{sub}$ that have $t_c$ `near' $t_{scan}$,
then the set $S_{t_{scan}}$ contains those $n$ fits.  These $n$ fits include
windows of varying lengths so that the windows themselves are likely associated
with different groups.  Now consider a given informative parameter $IP_l$ and
its underlying parameter $j$ (described in Sec.~\ref{sec:ip}) that has an
associated `good region', $Rg_{I,l}$.  Remember that this informative parameter
$IP_l$ has an associated group.  Count the number $p$ of the $n$ fits whose
lengths belong to the associated group of $IP_l$.  If more of the values of the
underlying parameter of $p$ lie within $Rg_{I,l}$ than outside of it, then
$IP_l$ belongs to Class I and, thus, the `answer' to the question of `belonging
to Class I' is $a = 1$.  If, on the other hand, more values lie outside the `good
region' $Rg_{I,l}$ than in it, the answer is $a = -1$.  If the same number of
values are inside and outside of $Rg_{I,l}$ then $a = 0$.  Also, if no members
of $S_{t_{scan}}$ belong to the associated group of $IP_l$ then $a = 0$.

To assist more in that understanding, let us have a look
at an example.  Assume that the informative parameter information tells us
parameter $m$ in Group 3 is the informative parameter $IP_{19}$ and $m \in [A,
B]$ is the `good region' $Rg_{I,l}$ of Class I.  We consider a single
$t_{scan}$ and find that there are two fits in $S_{t_{scan}}$ in this group
with parameter $m$ values of $m_1$ and $m_2$.  We determine the `answer' $a =
a_{IP_{19}}$ as follows:
\begin{itemize}
\item If $m_1, m_2 \in [A, B]$, we say that based on $IP_{19}$ (Group 3,
  parameter $m$) that fits near $t_{scan}$ belong to Class I. Mark this answer
  as $a_{IP_{19}} = 1$.
\item If $m_1 \in [A, B]$ and $m_2 \notin [A, B]$, we say that fits near
  $t_{scan}$ cannot be identified and so $a_{IP_{19}} = 0$.
\item If $m_1,m_2 \notin [A, B]$, fits near $t_{scan}$ belong to Class II
  and $a_{IP_{19}} = -1$.
\end{itemize}
More succinctly,
\begin{equation}
   a_{IP_{19}} = \left\{ \begin{array}{ll}
                          1 & \mbox{if $m_1,m_2 \in [A, B]$} \\
                          0 & \mbox{if $m_i \in [A, B], m_j \notin
                            [A, B], i \neq j, i,j \in \{1,2\}$}\\
                          -1 & \mbox{if $m_1,m_2 \notin [A, B]$}
                          \end{array}
                  \right.
\end{equation}

For each of the informative parameters, we get an answer $a$ that says that
fits near $t_{scan}$ belong to Class I or II (or cannot be determined). For a
total of $L$ informative parameters, we get a questionnaire $A$ of length $L$:
\begin{equation}
  A_{t_{scan}} = a_1a_2a_3...a_L, a_i \in \{-1,0,1\}
\end{equation}
Qualitatively, these questionnaires describe our judgement to whether $t_{scan}$ is a rebound or not. This judgement
depends on the observations of informative parameters.
\subsection{Traits}
\label{sub:trait}

The concept of a \emph{trait} is developed to describe the property
of the questionnaire for each $t_{scan}$. Each questionnaire can be
decomposed into a fixed number of traits if the length of
questionnaire is fixed.

From any questionnaire with length $L$, we generate a series of traits by the
following method.  Every trait is a series of 4 to 6 integers, $\tau = p, q, r,
(P,Q,R)$.  The first three terms $p, q$ and $r$ are simply integers.  The term
$(P, Q, R)$ represents a string of 1 to 3 integers.  We first describe $p, q$ and
$r$ and then the $(P, Q, R)$ term.

The integers $p, q$ and $r$ have limits: $p \in {1,2,\ldots,L}, q \in
{p,p+1,\ldots,L}, r \in {q,q+1,\ldots,L}$.  We select all the possible
combinations of bits from the questionnaire $A_{t_{scan}}$ with the condition
that each time the number of selected questions is at most 3.  We record the
numbers of the selected positions and sort them. The terms $p,q$ and $r$ are
selected position numbers and defined as follows:

\begin{itemize}
\item If only one position $i_1$ is selected: $r = q = p = i_1$
\item If two $i_1, i_2$ are selected: $p = i_1, r = q = i_2 (i_1 < i_2)$
\item If three $i_1,i_2,i_3$ are selected: $p = i_1, q = i_2, r = i_3 (i_1 <
  i_2 < i_3)$
\end{itemize}

The term $(P, Q, R)$ is defined as follows:
\begin{eqnarray}
  r = q = p,& &(P,Q,R) = a_p \\
  r = q, q \neq p,& &(P,Q,R) = a_p,a_q \\
  r \neq q, q \neq p,& &(P,Q,R) = a_p,a_q,a_r
\end{eqnarray}
As an example, A = (0,1,-1,-1) has traits in Table~\ref{tb:trait}.

For a questionnaire with length L, there are $3L + 3^2 {L \choose 2} + 3^3 {L \choose 3} $
possible traits. However, a single questionnaire has only $L + {L \choose 2}  + {L \choose 3}$
traits, because (P,Q,R) is defined by p,q and r.  In this example, there are 14
traits for questionnaire (0,1,-1,-1) and 174 total traits for all possible $L =
4$ questionnaires.

\subsection{Features}
\label{sec:features}

At the risk of being redundant, it is worth briefly summarizing again.  Until
now we have: $L$ informative parameters $IP_1,IP_2,\ldots,IP_L$ from 84
different parameters ($84 = 6~ \mathrm{parameters} \times 14~ \mathrm{Groups}$)
and a series of questionnaires $A_{t_{scan}}$ for each $t_{scan}$ from
$t_{cmin}$ to $t_{cmax}$ using set $S_{t_{scan}}$. These questionnaires depend
upon which subset $\Sigma_{sub}$ of fits is chosen.  Each questionnaire has a
sequence of traits that describe the property of this questionnaire in a short and clear way.
Now we generate \emph{features} for both classes.

Recall that the subset of fits $\Sigma_{feature}$ that we use here is that which contains
all fits which have a critical time $t_c$ earlier than $t_p =$ 1975-01-01,
$\Sigma_{feature} = \{ f \mid t_{c,f} < t_p \}$.
By imposing that $t_2$ and $t_{c,f}$ are both smaller than $t_p$, we do not use any future
information.  Considering the boundary condition of critical times in
Eq.~(\ref{eq:rangetc}), the end time of a certain fit $t_2$ is less than or equal to
$t_c$. Additionally, we select only those critical times such that $t_{c,f} < t_p, \forall f \in
\Sigma_{feature}$.

Assume that there are two sets of traits $T_I$ and $T_{II}$ corresponding to
Class I and Class II, respectively.  Scan day by day the date $t$ from the smallest $t_c$
in $\Sigma_{feature}$ until $t_p$. If $t$ is near a rebound (using
the same $D = 10$ day criterion as before), then all traits generated by
questionnaire $A_t$ belong to $T_I$. Otherwise, all traits generated by $A_t$
belong to $T_{II}$.

Count the frequencies of a single trait $\tau$ in $T_I$ and $T_{II}$. If $\tau$
is in $T_I$ for more than $\alpha$ times and in $T_{II}$ for less than $\beta$
times, then we call this trait $\tau$ a \emph{feature} $F_I$ of Class I.
Similarly, if $\tau$ is in $T_{I}$ for less than $\alpha$ times and in
$T_{II}$ for more than $\beta$ times, then we call $\tau$ a \emph{feature} $F_{II}$ of
Class II.  The pair $(\alpha, \beta)$ is defined as a \emph{feature
  qualification}.  We will vary this qualification to optimize the back tests
and predictions.

\subsection{Rebound alarm index}

The final piece in our methodology is to define a \emph{rebound alarm index}
that will be used in the forward testing to `predict' rebounds.  Two types of
rebound alarm index are developed.  One is for the back tests before
1975-01-01, as we have already used the information before this time to
generate informative parameters and features.  The other alarm index is for the
prediction tests.  We generate this prediction rebound alarm index using only
the information before a certain time and then try to predict rebounds in the
`future' beyond that time.

\section{Back testing}
\label{sub:back}

\subsection{Features of learning set}
\label{sec:features-learning}

Recall that a \emph{feature} is a trait which frequently appears in
one class but rarely in the other class. Features are associated
with feature qualification pairs $(\alpha, \beta)$. Using all the
fits from subset $\Sigma_{feature}$ found in
Sec.~\ref{sec:features}, we generate the questionnaires for each day
in the learning set, i.e., the fits with $t_c$ before 1975-01-01.
Take all traits from the questionnaire $A_t$ for a particular day
$t$ and compare them with features $F_I$ and $F_{II}$. The number of
traits in $F_I$ and $F_{II}$ are called $\nu_{t,I}$ and
$\nu_{t,II}$. Then we define:
\begin{equation}
  RI_t = \left \{ \begin{array}{ll}
      \frac{\nu_{t,I}}{\nu_{t,I}+\nu_{t,II}}& \mbox{if
      $\nu_{t,I}+\nu_{t,II} \geq 0$}\\
      0& \mbox{if $\nu_{t,I}+\nu_{t,II} = 0$}
                  \end{array}
                  \right.
\end{equation}
From the definition, we can see that $RI_t \in [0,1]$. If $RI_t$ is high,
then we expect that this day has a high probability that the rebound will start.

We choose feature qualification pair (10, 200) here, meaning that a
certain trait must appear in trait Class I at least 11 times
\emph{and} must appear in trait Class II less than 200 times.  If
so, then we say that this trait is \emph{a feature of Class I}.  If,
on the other hand, the trait appears 10 times or less in Class I
\emph{or} appears 200 times or more in Class II, then this trait is
\emph{a feature of Class II}. The result of this feature
qualification is shown in Figure \ref{fg:f2212}.  Note that the
choice (10, 200)  is somewhat arbitrary and does not constitute an
in-sample optimization on our part. This can be checked from the
error diagrams presented below, which scan these numbers: one can
observe in particular that the pair (10, 200) does not give the best
performance. We have also investigated the impact of changing other
parameters and find a strong robustness.

With this feature qualification, the rebound alarm index can
distinguish rebounds with high significance. If the first number
$\alpha$ is too big and the second number $\beta$ is too small, then
the total number of Class I features will be very small and the
number of features in Class II will be large.  This makes the
rebound alarm index always close to 0.  In contrast, if $\alpha$ is
too small and $\beta$ is too large, the rebound alarm index will
often be close to 1. Neither of these cases, then, is qualified to
be a good rebound alarm index to indicate the start of the next
rebound. However, the absolute values of feature qualification pair
are not very sensitive within a large range. Only the ratio
$\alpha/\beta$ plays an important role. Figures \ref{fg:22p} -
\ref{fg:33b} show that varying $\alpha$ and $\beta$ in the intervals
$10 \leq \alpha \leq 20$ and $200 \leq \beta \leq 1000$ does not
change the result much. For the sake of conciseness, only the
rebound alarm index of feature qualification pair (10, 200) is shown
in this paper.

\subsection{Predictions}

Once we generate the Class I and II features of the learning set for values of
$t_c$ before $t_p$ (Jan. 1, 1975), we then use these features to generate the
predictions on the data after $t_p$.  Recall that the windows that we fit are
defined such that the end time $t_2$ increases 50 days from one window to the
next.  Also note that all predictions made on days between these 50 days will
be the same because there is no new fit information between, say, $t_{2}^n$ and
$t_2^{n - 1}$.

Assume that we make a prediction at time $t$:
\begin{equation}
  t \in (t_2, t_2 + 50], ~~t > t_p
\end{equation}
Then the fits set $\Sigma_{t_2} = \{f \mid t_{2,f} \leq t_2 \}$ is made using
the past information before prediction day $t$.  We use $\Sigma_{t_2}$ as the
subset $\Sigma_{sub}$ mentioned in Sec.~\ref{sub:question} to generate the
questionnaire on day $t$ and the traits for this questionnaire.  Comparing
these traits with features $F_I$ and $F_{II}$ allows us to generate a rebound
alarm index $RI_t$ using the same method as described in Sec.~\ref{sec:features-learning}.

Using this technique, the prediction day $t_2$ is scanned from
1975-01-01 until 2009-07-22  in steps of 50 days. We then construct
the time series of the \emph{rebound alarm index}
over this period and with this resolution of 50 days.
The comparison of this rebound alarm
index with the historical financial index (Figure~\ref{fg:p2212}) shows
a good correlation, but there are
also some false positive alarms (1977, 1998, 2006), as well as some false
negative missed rebounds (1990). Many false positive alarms such as
in 1998 and 2006 are actually associated with rebounds. But these rebounds
have smaller amplitudes than our qualifying threshold targets.
Concerning the false negative (missed rebound) in 1990, the explanation is probably that
the historical prices preceeding this rebound does not follow the JLS model specification.
Rebounds may result from several mechanisms and the JLS model only provides one of them,
arguably the most important. Overall, the predictability of the rebound alarm
index shown in Figure~\ref{fg:p2212}, as well as the relative cost of
the two types of errors (false positives and false negatives) can be quantified
systematically, as explained in the following sections. The major conclusion
is that the rebound alarm index has a prediction skill much better than luck,
as quantified by error diagrams.

\subsection{Error Diagram}
\label{sec:error}

We have qualitatively seen that the feature qualifications method using back
testing and forward prediction can generate a rebound alarm index that seems to
detect and predict well observed rebounds in the S\&P 500 index. We now
quantify the quality of these predictions with the use of error diagrams \cite{Molchan1,Molchan2}.
We create an error diagram for predictions after 1975-01-01 with a
certain feature qualification in the following way:
\begin{enumerate}
\item Count the number of rebounds after 1975-01-01
as defined in section \ref{defreboundh2ysec} and expression (\ref{defreboundh2y}).
There are 9 rebounds.
\item Take the rebound alarm index time series (after 1975-01-01) and sort the set
of all alarm index values in decreasing order. There are 12,600 points in this series
and the sorting operation delivers a list of 12,600 index values, from the largest
to the smallest one.
\item  The largest value of this sorted series defines the first
  threshold.
\item Using this threshold, we declare that an alarm starts on the first day
  that the unsorted rebound alarm index time series exceeds this threshold.
  The duration of this alarm $D_a$ is set to 41 days, since the longest
  distance between a rebound and the day with index greater than the threshold
  is 20 days.  Then, a prediction is deemed successful when a rebound
  falls inside that window of 41 days.
\item If there are no successful predictions at this threshold, move the
  threshold down to the next value in the sorted series of alarm index.
\item Once a rebound is predicted with a new value of the threshold, count the
  ratio of unpredicted rebounds (unpredicted rebounds / total rebounds in set)
  and the ratio of alarms used (duration of alarm period / 12,600 prediction
  days). Mark this as a single point in the error diagram.
\end{enumerate}
In this way, we will mark 9 points in the error diagram for the 9 rebounds.

The aim of using such an error diagram in general is to show that a given
prediction scheme performs better than random.  A random prediction follows the
line $y = 1 - x$ in the error diagram. A set of points below this line
indicates that the prediction is better than randomly choosing alarms. The
prediction is seen to improve as more error diagram points are found near the
origin (0, 0). The advantage of error diagrams is to avoid discussing how
different observers would rate the quality of predictions in terms of
the relative importance of avoiding the occurrence of
false positive alarms and of false negative missed rebounds.
By presenting the full error diagram, we thus sample all possible
preferences and the unique criterion is that the error diagram curve
be shown to be statistically significantly below the anti-diagonal $y = 1 - x$.

In Figure \ref{fg:22p}, we show error diagrams for different feature qualification pairs $(\alpha, \beta)$.
Note the 9 points representing
the 9 rebounds in the prediction set.  We also plot the 11 points of the error
diagrams for the learning set in Figure \ref{fg:22b}.

As a different test of the quality of this pattern recognition procedure, we
repeated the entire process but with a rebound now defined as the minimum price
within a window of $2 \times 365$ days\footnote{seven rebounds
in the back tests before 1975.1.1: 1953-09-14; 1957-10-22; 1960-10-25;
1962-06-26; 1966-10-07; 1970-05-26; 1974-10-03,
 and six rebounds after 1975.1.1
 in the prediction range: 1978-03-06; 1982-08-12; 1987-12-04; 1990-10-11;
 2002-10-09; 2009-03-09.} instead of $2 \times 200$ days, as
before. These results are shown in Figures~\ref{fg:33p}-\ref{fg:33b}.

\subsection{Bayesian inference}

Given a value of the \emph{predictive} rebound alarm index, we can also use the
\emph{historical} rebound alarm index combined with Bayesian inference to calculate the
probability that this value of the rebound alarm index will actually be followed by a rebound.  We use
predictions near the end of November, 2008 as an example.  From
Figure~\ref{fg:p2212}, we can see there is a strong rebound signal in that
period.  We determine if this is a true rebound signal by the following
method:
\begin{enumerate}
\item Find the highest rebound alarm index $Lv$ around the end of November
  2008.
\item Calculate $D_{total}$, the number of days in the interval from 1975-01-01
  until the end of the prediction set, 2009-07-22.
\item Calculate $D_{Lv}$, the number of days which have a rebound alarm index
  greater than or equal to $Lv$.
\item The probability that the rebound alarm index is higher than $Lv$ is estimated by
\begin{equation}
  P(RI \geq Lv) = \frac{D_{Lv}}{D_{total}}
\end{equation}
\item The probability of a day being near the bottom of a rebound is
estimated as the number
of days near real rebounds over the total number of days in the predicting set:
\begin{equation}
  P(rebound) = \frac{D_{rw} N_{rebound}}{D_{total}},
\end{equation}
where $N_{rebound}$ is the number of rebounds we can detect after 1975-01-01
and $D_{rw}$ is the rebound width, i.e. the number of days near the real
rebound in which we can say that this is a successful prediction.  For example,
if we say that the prediction is good when the predicted rebound time and real
rebound time are within 10 days of each other, then the rebound width $D_{rw} =
10 \times 2 + 1 = 21$.
\item The probability that the neighbor of a rebound has a rebound alarm index
  larger than $Lv$ is estimated as
\begin{equation}
  P(RI \geq Lv | rebound) = \frac{N_0}{N_{rebound}}
\end{equation}
where $N_0$ is the number of rebounds in which
\begin{equation}
\sup_{|d-rebound| \leq 20} RI_d \geq Lv.
\end{equation}
\item Given that the rebound alarm index is higher than $Lv$, the probability
  that the rebound will happen in this period is given by Bayesian inference:
\begin{equation}
  P(rebound | RI \geq Lv) = \frac{P(rebound)\times P(RI \geq Lv | rebound)}{P(RI \geq Lv)}
\end{equation}
\end{enumerate}

Averaging $P(rebound | RI \geq Lv)$ for all the different feature
qualifications gives the probability that the end of November 2008 is a rebound
as 0.044. By comparing with observations, we see that this period is not a
rebound.  We obtain a similar result by increasing the definition of rebound from
200 days before and after a local minimum to 365 days, yielding a probability
of 0.060.

When we \emph{decrease} the definition to 100 days, the probability
that this period is a rebound jumps to 0.597.  The reason for this sudden jump
is shown in Figure~\ref{fg:bn08} where we see the index around this period and the
S\&P 500 index value. From the figure, we find that this period is a local
minimum within 100 days, not more. This is consistent with what Bayesian
inference tells us. However, we have to address that the more obvious rebound
in March 2009 is missing in our rebound alarm index. Technically, one can easily find that
this is because the end of crash is not consistent with the beginning of rebound in this special period.

In this case, we then test all the days after 1985-01-01 systematically by Bayesian inference
using only prediction data (rebound alarm index) after 1975-01-01. To show that
the probability that $RI \geq Lv$ is stable, we cannot start Bayesian inference
too close to the initial predictions so we choose 1985-01-01 as the beginning
time.  We have 5 `bottoms' (troughs) after this date, using the definition of a
minimum within $\pm 200$ days.

For a given day $d$ after 1985-01-01, we know all values of the rebound alarm
index from 1975-01-01 to that day.  Then we use this index and historical data
of the asset price time series in this time range to calculate the probability
that $d$ is the bottom of the trough, given that the rebound alarm index is larger
than $Lv$, where $Lv$ is defined as
\begin{equation}
  Lv = \sup_{d-t<50} RI_t
\end{equation}

To simplify the test, we only consider the case of feature qualification pair
(10, 200), meaning that the trait is a feature of Class I only if it shows in
Class I more than 10 times and in Class II less than 200 times.
Figure \ref{fg:bayes} shows that the actual rebounds occur near
the local highest probability of rebound calculated by Bayesian inference.
This figure also illustrates the existence of false positive alarms, i.e., large
peaks of the probability not associated with rebounds that we have characterized
unambiguously at the time scale of $\pm 200$ days.

\section{Trading strategy}

In order to determine if the predictive power of our method provides a genuine and useful
information gain, it is necessary to estimate the excess return it could generate.
The excess return is the real return minus the risk free rate
transformed from annualized to the duration of this period. The annualized 3-month US treasury bill rate is
used as the risk free rate in this paper.
We thus develop a trading strategy based on the rebound alarm index as follows.  When the
rebound alarm index rises higher than a threshold value $Th$, then with a lag of $Os$ days,
we buy the asset.  This entry strategy is complemented by the following exit strategy.
When the rebound alarm index goes below $Th$, we still hold the
stock for another $Hp$ days, with one exception. Consider the case that the
rebound alarm index goes below $Th$ at time $t_1$ and then rises above $Th$
again at time $t_2$.  If $t_2 - t_1$ is smaller than the holding period $Hp$,
then we continue to hold the stock until the next time when the rebound alarm index remains
below $Th$ for $Hp$ days.

The performance of this strategy for some fixed values of the parameters is
compared with random strategies, which share all the properties except
for the timing of entries and exits determined by the rebound alarm index
and the above rules. The random strategies
consist in buying and selling at random times, with the constraint that the
total holding period (sum of the holding days over all trades in a given strategy)
is the same as in the realized strategy that we test.
Implementing 1000 times these constrained random strategies with different
random number realizations provide
the confidence intervals to assess whether the performance of our strategy
can be attributed to real skill or just to luck.

Results of this comparison are shown in Table~\ref{tb:performance} for two sets of
parameter values.  The p-value
is a measure of the strategies' performance, calculated as the fraction of
corresponding random strategies that are better than or equal to our
strategies.  The lower the p-value is, the better the strategy is compared to
the random portfolios.  We see that all of our strategies'  cumulative excess returns are among the top 5-6\%
out of 1000 corresponding random strategies' cumulative excess returns.  Box plots for each of the
strategies are also presented in Figures~\ref{fg:bp1}-\ref{fg:bp2}.

The cumulative returns as well as the cumulative excess returns obtained with the two strategies
as a function of time are shown in  Figures~\ref{fg:wt1}-\ref{fg:wt2}.
These results suggest that these two strategies would provide
significant positive excess return. Of course, the performance obtained here
are smaller than the naive buy-and-hold strategy, consisting in buying at the
beginning of the period and just holding the position. The comparison
with the buy-and-hold strategy would be however unfair
as our strategy is quite seldom invested in the market.
Our goal here is not to do better than any other strategy
but to determine the statistical significance of a specific signal. For this, the
correct method is to compare with random strategies that are invested
in the market the same fraction of time. It is obvious that we could improve
the performance of our strategy by combining
the alarm indexes of bubbles and of negative bubbles, for instance, but this
is not the goal here.

We also provide the Sharpe ratio as
a measure of the excess return (or risk premium) per unit of risk. We define it per trade as follows
\begin{equation}
    S = \frac{E[R - R_f]}{\sigma}
\end{equation}
where $R$ is the return of a trade, $R_f$ is the risk free rate (we use the 3-month US treasury bill rate)
transformed from annualized to the duration of this trade given in Table~\ref{tb:performance}
and $\sigma$ is the standard deviation of the returns per trade. The higher the Sharpe ratio is,
the higher the excess return under the same risk.

The bias ratio is defined as the number of trades with a positive return within one standard deviation divided by one plus the number of
trades which have a negative return within one standard deviation:
\begin{equation}
    BR = \frac{\# \{r | r \in [0, \sigma]\}}{1 + \# \{r | r \in [-\sigma, 0)\}}
    \label{eq:br}
\end{equation}
In Eq.~(\ref{eq:br}), $r$ is the excess return of a trade and $\sigma$ is the standard deviation of the excess returns.
This ratio detects valuation bias.

To see the performance of our strategies, we also check all the possible
random trades with a holding period equals to the average duration of our strategies,
namely 25 days and 17 days for strategy I and II respectively.
The average Sharpe and bias ratios of these random trades are shown in Table~\ref{tb:performance}.
Both Sharpe and bias ratios of our strategies are greater than those of the random trades,
confirming that our strategies deliver a larger excess return with a stronger
asymmetry towards positive versus negative returns.

As another test, we select randomly the same number of random trades as in our strategies,
making sure that there is no overlap between the selected trades. We calculate the
Sharpe and bias ratios for these random trades. Repeating this random comparative selection
1000 times provides us with p-values for the Sharpe ratio and for bias ratio of our strategies.
The results are presented in Table~\ref{tb:performance}.
All the p-values are found quite small, confirming that our strategies perform well.

\section{Conclusion}

We have developed a systematic method to detect rebounds in financial markets
using ``negative bubbles,'' defined as the symmetric of
bubbles with respect to a horizontal line, i.e., downward accelerated
price drops. The aggregation of thousands of calibrations in running windows
of the negative bubble model on financial data has been performed using a general pattern
recognition method, leading to the calculation of a rebound alarm index.
Performance metrics have been presented in the form of
error diagrams, of Bayesian inference to determine the probability of rebounds
and of trading strategies derived from the rebound alarm index dynamics.
These different measures suggest that the rebound alarm index provides
genuine information and suggest predictive ability. The implemented
trading strategies outperform randomly chosen
portfolios constructed with the same statistical characteristics.
This suggests that financial markets may be characterized by transient
positive feedbacks leading to accelerated drawdowns,
which develop similarly to but as mirror images
of upward accelerating bubbles. Our key result is that these negative bubbles
have been shown to be predictably associated with
large rebounds or rallies.

In summary, we have expanded the evidence for the possibility to
diagnose bubbles before they terminate \cite{BFE-FCO10}, by adding
the phenomenology and modeling of ``negative bubbles'' and their
anticipatory relationship with rebounds. The present paper
contributes to improving our understanding of the most dramatic
anomalies exhibited by financial markets in the form of
extraordinary deviations from fundamental prices (both upward and
downward) and of extreme crashes and rallies. Our results suggest a
common underlying origin to both positive and negative bubbles in
the form of transient positive feedbacks leading to identifiable and
reproducible faster-than-exponential price signatures.

\newpage
\section{List of Symbols}

\begin{table}[H]
\caption{\label{tb:symbol} List of symbols}
\begin{tabular}{cl}
\hline\hline
$h(t)$&hazard rate\\
$p(t)$&stock price\\
$A,B,C$&linear parameters of the JLS model\\
$t_c$&critical time in the JLS model at which the bubble ends\\
$m$& exponent parameter in the JLS model\\
$\omega$&frequency parameter in the JLS model\\
$\phi$&phase parameter in the JLS model\\
$b$&parameter controlling the positivity of the hazard rate in the JLS model \\
$Rbd$&rebound time\\
$C_I$& set of Class I fits\\
$C_{II}$& set of Class II fits\\
$G_i$&set of Group i fits\\
$IP$&informative parameter\\
$A$&questionnaire\\
$\tau$&trait\\
$(\alpha,\beta)$&feature qualification pair\\
$RI$&rebound alarm index\\
$Lv$&highest rebound alarm index around a certain time\\
$Th$&threshold value for the trading strategy\\
$Os$&offset  for the trading strategy\\
$Hp$&holding period  for the trading strategy\\
$S$&Sharpe ratio\\
$R_f$&risk free rate\\
$r, R-R_f$& excess return of a trade\\
$BR$&bias ratio\\
$\#$&number of a set\\
\hline\hline
\end{tabular}
\end{table}

\begin{table}
\caption{\label{tb:trait}Traits for series A = (0,1,-1,-1)}
\begin{tabular}{llll}
\hline\hline

p&q&r&(P,Q,R)\\\hline
1&1&1&0\\
1&2&2&0,1\\
1&2&3&0,1,-1\\
1&2&4&0,1,-1\\
1&3&3&0,-1\\
1&3&4&0,-1,-1\\
1&4&4&0,-1\\
2&2&2&1\\
2&3&3&1,-1\\
2&3&4&1,-1,-1\\
2&4&4&1,-1\\
3&3&3&-1\\
3&4&4&-1,-1\\
4&4&4&-1\\\hline\hline
\end{tabular}
\end{table}

\begin{table}
\caption{\label{tb:performance}Performances of two strategies: Strategy I ($Th=0.2,Os=10,Hp=10$) and Strategy II ($Th=0.7,Os=30,Hp=10$).}
\begin{tabular}{lll}
\hline\hline
 &Strategy I&Strategy II\\
 \hline
Threshold ~$Th$ &0.2&0.7\\
Offset ~$Os$ &10&30\\
Holding period ~$Hp$ &10&10\\
Number of trades &77&38\\
Success rate & &\\
(fraction of trades with positive return)&66.2\%&65.8\%\\
Total holding days&1894 days&656 days\\
Fraction of time when invested &15.0\%&5.2\%\\
Cumulated log-return&95\%&45\%\\
cumulated excess log-return&67\%&35\%\\
 Average return per trade&1.23\%&1.19\%\\
 Average trade duration &24.60 days&17.26 days\\
 $p$-value of cumulative excess return&0.055&0.058\\
 Sharpe ratio per trade &0.247&0.359\\
 Sharpe ratio of random trades & & \\
 (holding period equals average trade duration)&0.025&0.021\\
 $p$-value of Sharpe ratio&0.043&0.036\\
 Bias ratio &1.70&1.36\\
 Bias ratio of random trades & &\\
 (holding period equals average trade duration)&1.27&1.25\\
 $p$-value of bias ratio&0.105&0.309\\
 \hline\hline
\end{tabular}
\end{table}

\clearpage
%\section{Bibliography}
%\bibliography{ijf.bib}

\section*{References}
%\bibliographystyle{elsart-harv}
%\bibliography{njpref.bib}

\clearpage

\begin{figure}[htbp]
\centering
\includegraphics[width=0.9\textwidth]{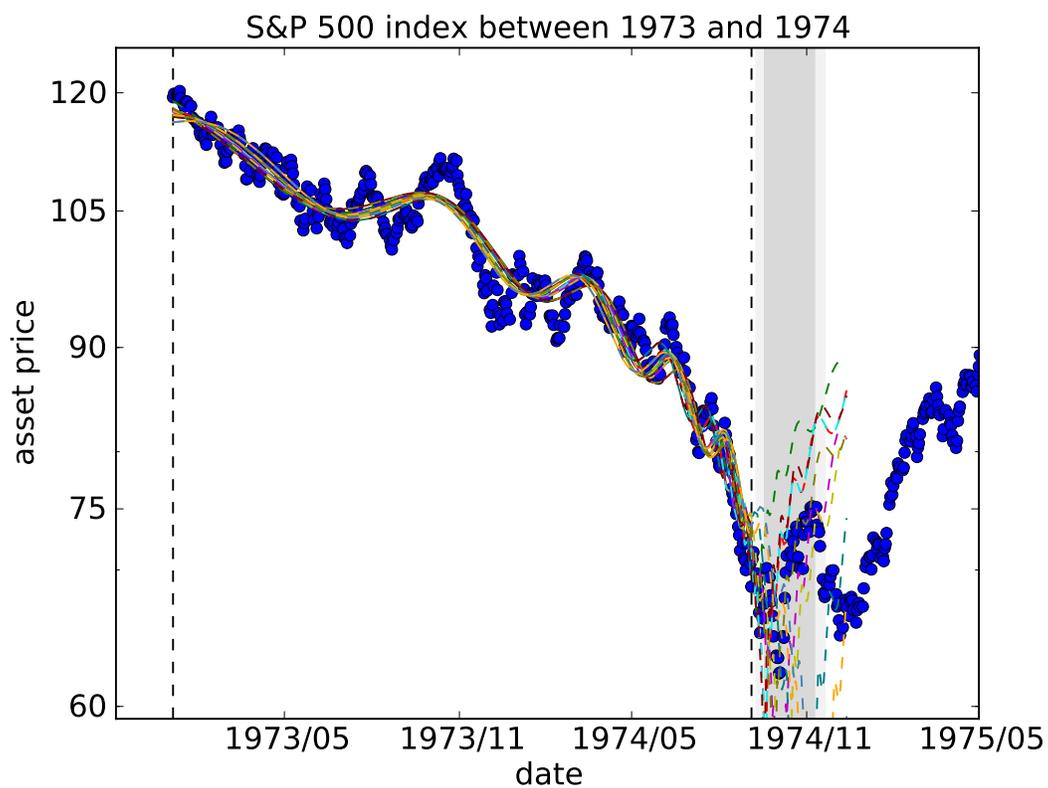}
\includegraphics[width=0.9\textwidth]{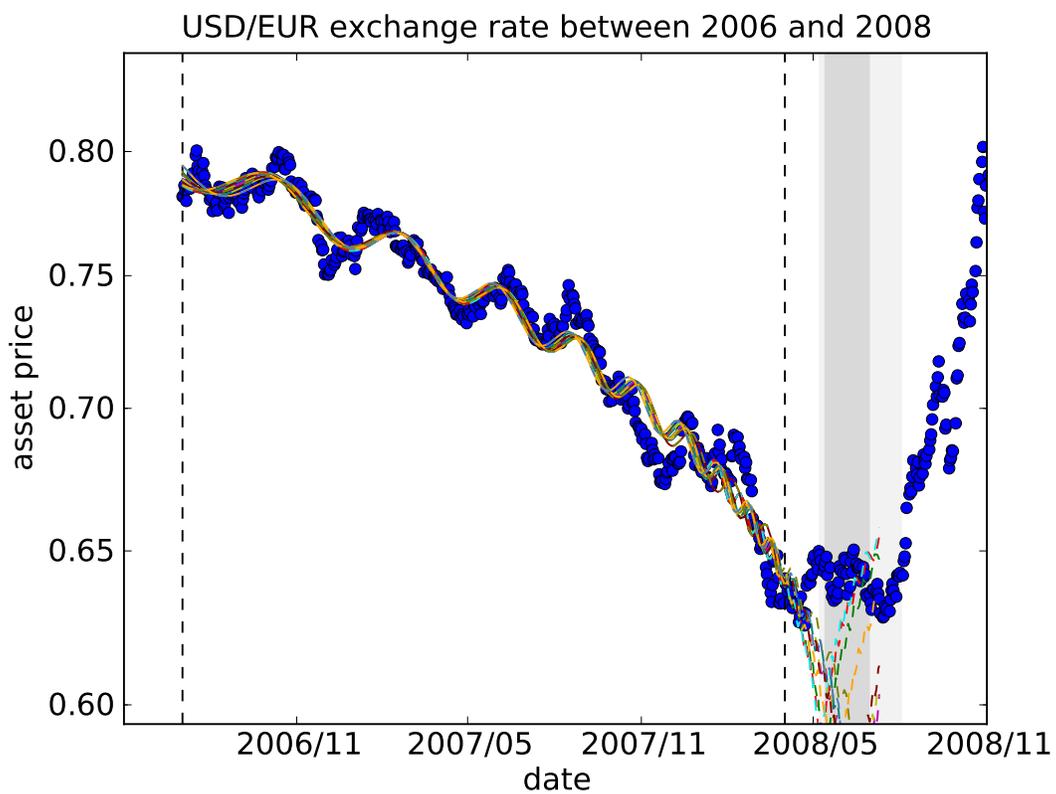}
\caption{Upper panel: Significant drawdown of nearly $50\%$ from
1973-01-01 to 1974-10-01 (time window delineated by the two black
dashed vertical lines) with very clear log-periodic oscillations,
followed by a strong positive rebound. The best fits from taboo
search are used to form a 90\% confidence interval for the critical
time $t_c$ shown by the light shadow area. The dark shadow area
corresponds to the 20-80 quantiles region of the predicted rebounds.
Lower panel: the same phenomenon is observed in foreign exchange
market. The plot shows the fitted results for USD/EUR change rate
from 2006-07-01 to 2008-04-01. The USD/EUR change rate performed a
significant drawdown with very clear log-periodic oscillations,
followed by a strong positive rebound.} \label{fg:lppl}
\end{figure}

\clearpage

\begin{figure}[htbp]
\centering
\includegraphics[width=\textwidth]{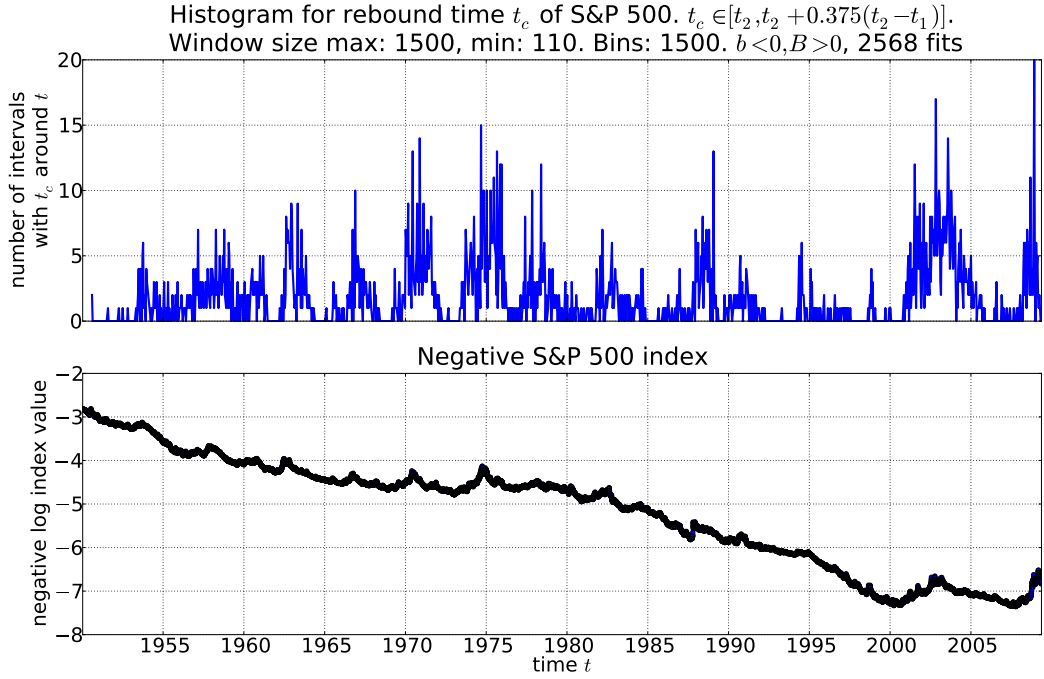}
\caption{(upper) Histogram of the critical times $t_c$ over the set of 2,568 time intervals
for which negative bubbles are detected by the condition that
the fits of $\ln p(t)$ by expression  (\ref{eq:lppl}) satisfy condition (\protect\ref{eq:rb}).
(lower) Plot of $- \ln p(t)$ versus time for the S\&P 500 index. Note that peaks in this figure correspond to valleys in actual price.}
\label{fg:histo}
\end{figure}

\clearpage

\begin{figure}[htbp]
\centering
\includegraphics[width=\textwidth]{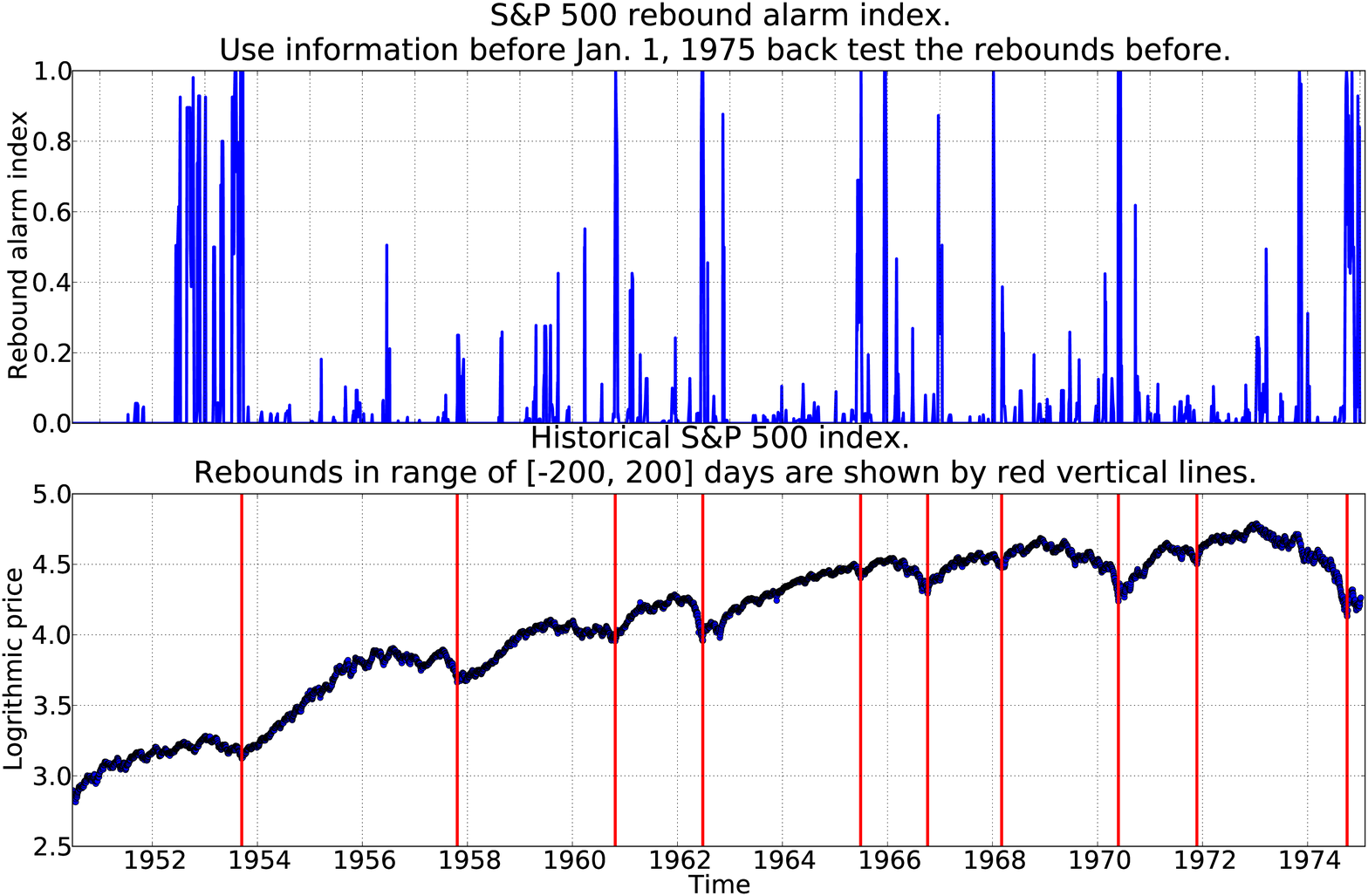}
\caption{Rebound alarm index and log-price of the S\&P 500 Index for
the learning set, where $t_2$ and $t_c$ are both before Jan. 1,
1975. (upper) Rebound alarm index for the learning set using feature
qualification pair $(10, 200)$. The rebound alarm index is in the
range $[0,1]$. The higher the rebound alarm index, the more likely
is the occurrence of a rebound. (lower) Plot of $\ln p(t)$ versus
time of S\&P Index. Red vertical lines indicate rebounds defined by
local minima within plus and minus 200 days around them. Note that
these rebounds are the historical ``change of regime'' rather than
only the jump-like reversals. The jump-like reversals, 1972, 1974 as
examples, are included in these rebounds. They are located near
clusters of high values of the rebound alarm index of the upper
figure.} \label{fg:f2212}
\end{figure}

\clearpage

\begin{figure}[htbp]
\centering
\includegraphics[width=\textwidth]{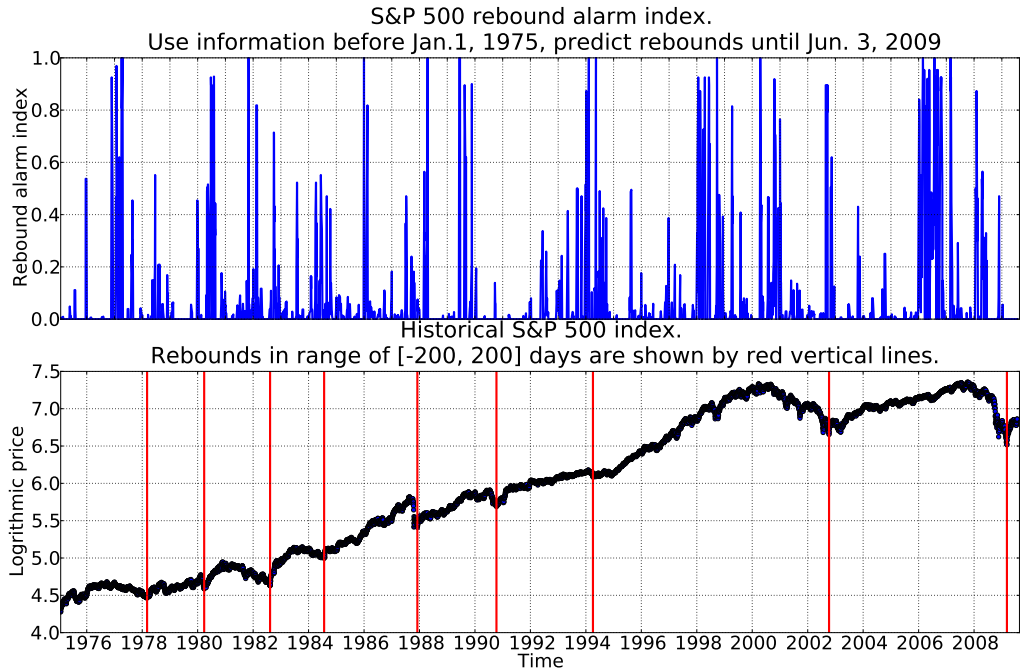}
\caption{Rebound alarm index and log-price of S\&P 500 Index for the predicting set after Jan. 1, 1975.
(upper) Rebound alarm index for predicting set using feature qualification pair $(10, 200)$. The rebound alarm index is in the
  range [0,1]. The higher the rebound alarm index, the more likely is the
  occurrence of a rebound. (lower) Plot of $\ln p(t)$ versus time of the S\&P Index. Red vertical lines indicate rebounds defined by local minima within in plus and minus 200 days.
  They are located near clusters of high values of the rebound alarm index of the upper figure.} \label{fg:p2212}
\end{figure}

\clearpage

\begin{figure}[htbp]
\centering
\includegraphics[width=\textwidth]{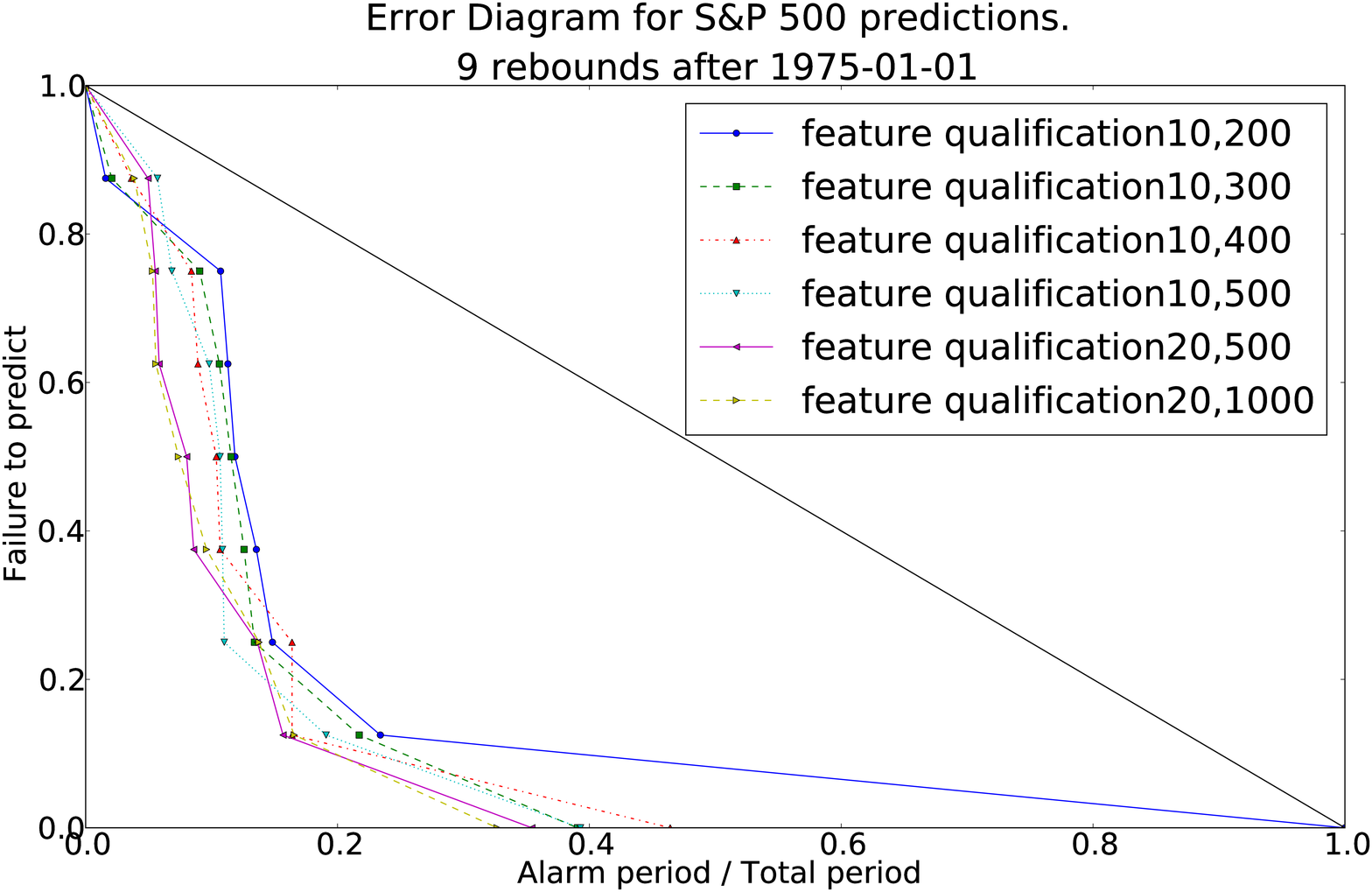}
\caption{Error diagram for predictions after Jan. 1, 1975 with different types
  of feature qualifications. Feature qualification $\alpha, \beta$ means that, if the
  occurrence of a certain trait in Class I is larger than $\alpha$ and less
  than $\beta$, then we call this trait a feature of Class I and vice
  versa. See text for more information.}
\label{fg:22p}
\end{figure}

\begin{figure}[htbp]
\centering
\includegraphics[width=\textwidth]{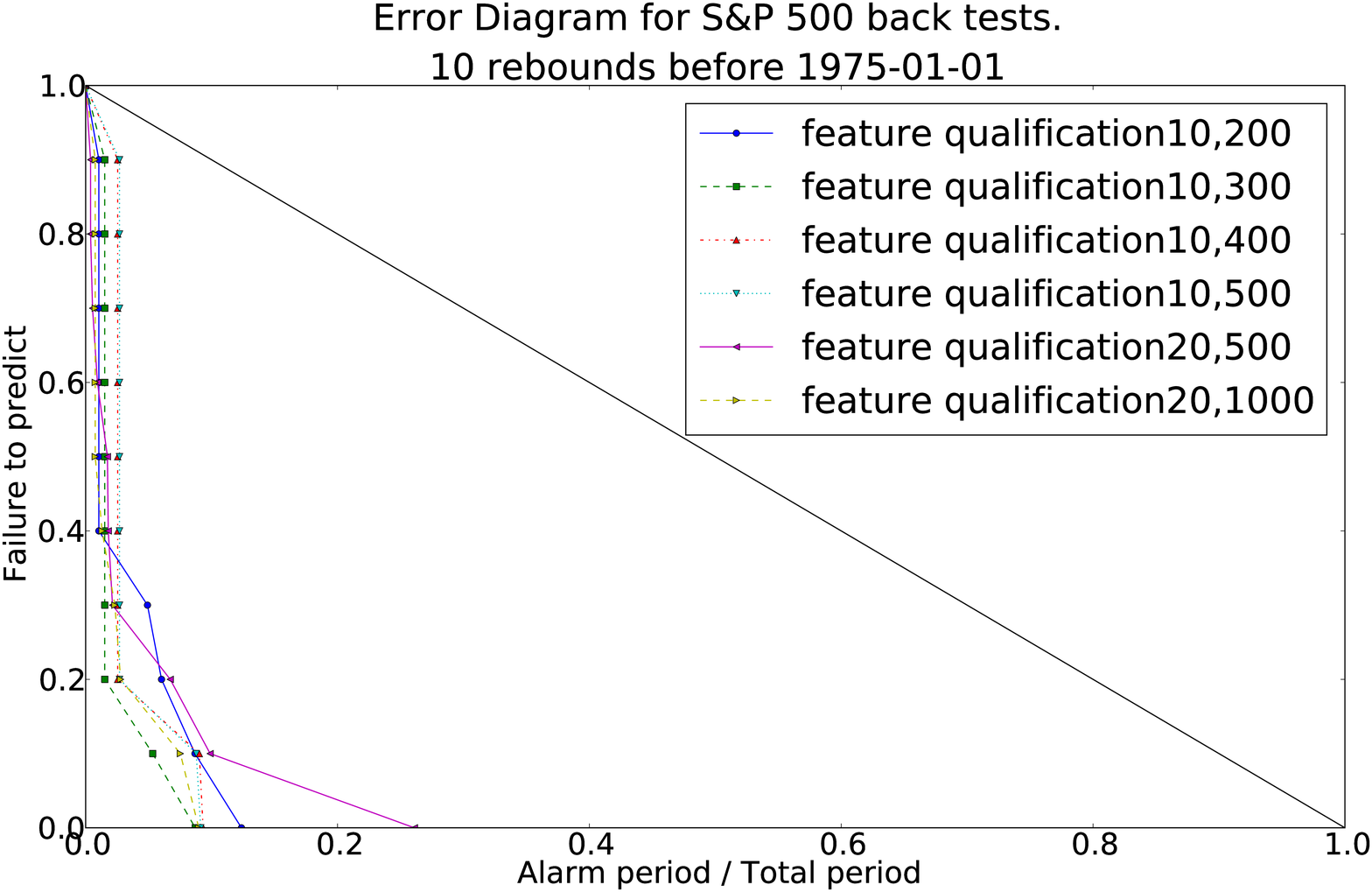}
\caption{Same as Figure \protect\ref{fg:22p} but for the learning set before Jan. 1, 1975.}
\label{fg:22b}
\end{figure}

\begin{figure}[htbp]
\centering
\includegraphics[width=\textwidth]{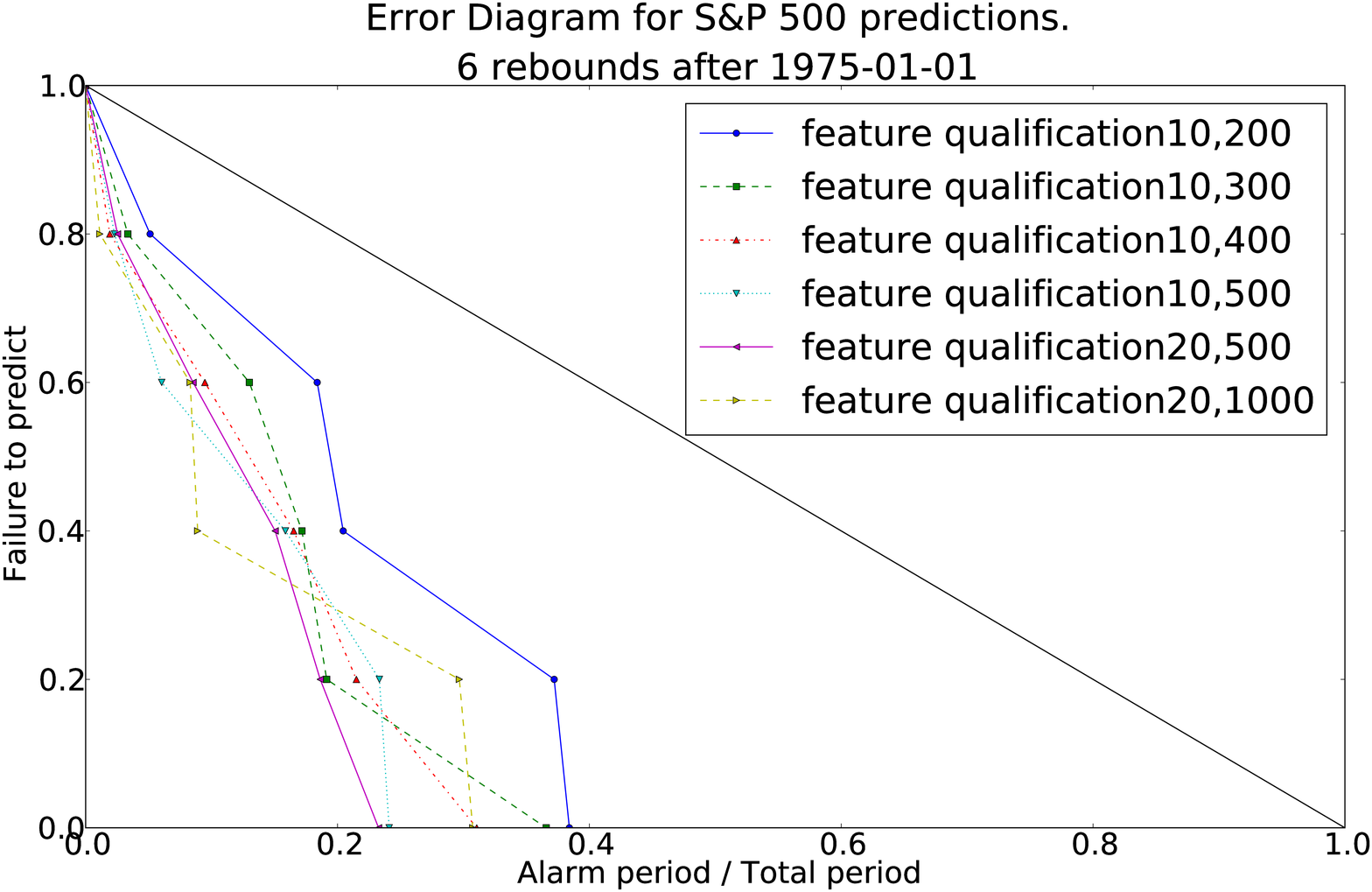}
\caption{Same as Figure \protect\ref{fg:22p} but with the different definition of
a rebound determined as the day with the smallest price within the 365 days
before it and the 365 days after it.}
\label{fg:33p}
\end{figure}

\begin{figure}[htbp]
\centering
\includegraphics[width=\textwidth]{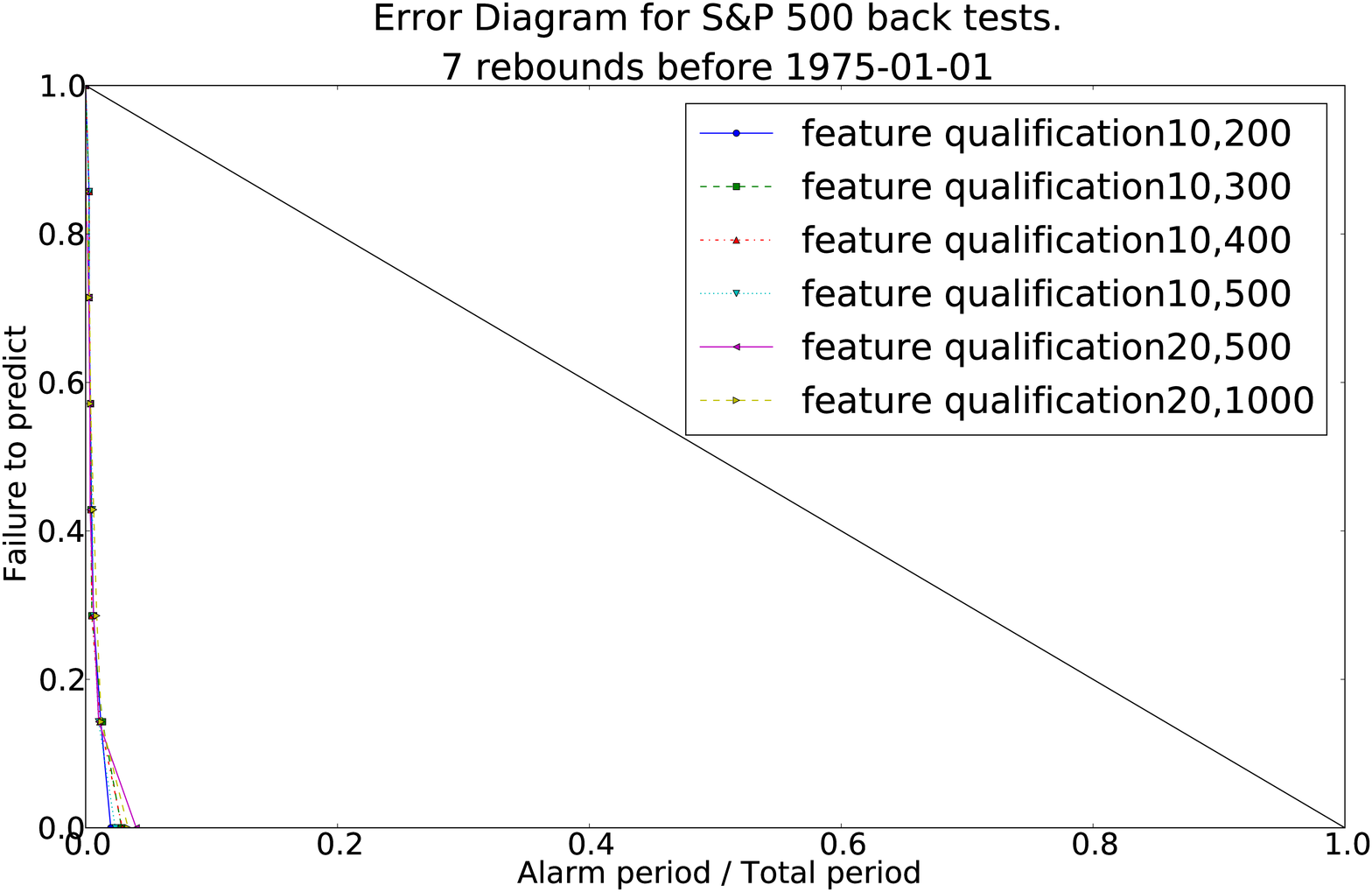}
\caption{Same as Figure \protect\ref{fg:33p} for the learning set before Jan. 1, 1975.}
\label{fg:33b}
\end{figure}

\clearpage

\begin{figure}[htbp]
\centering
\includegraphics[width=\textwidth]{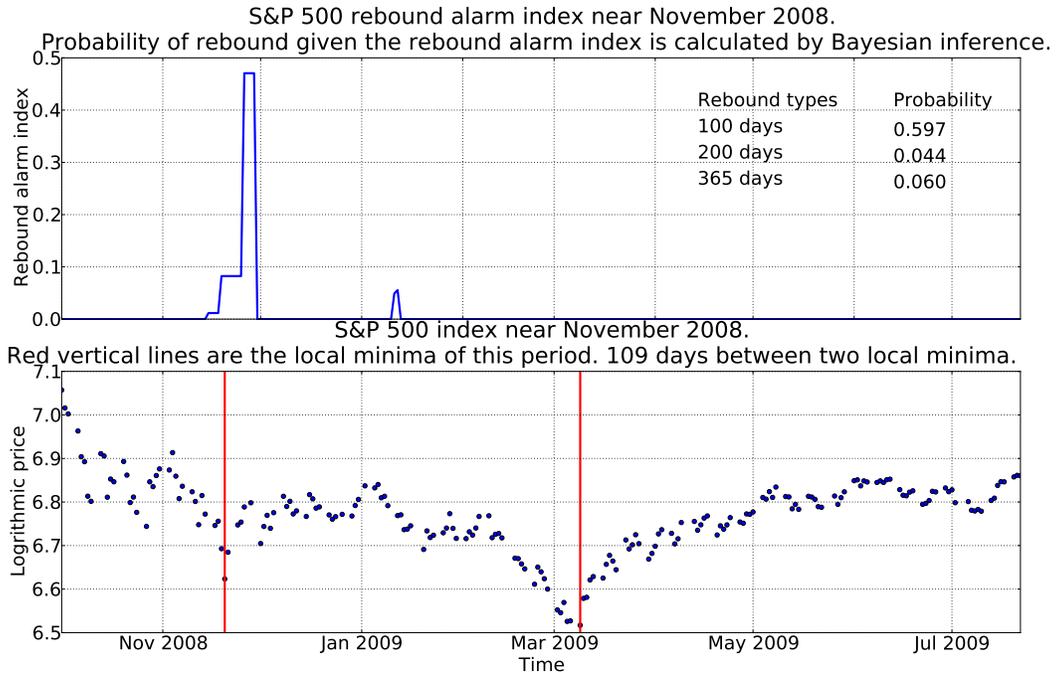}
\caption{Rebound alarm index and market price near and after November 2008.}
\label{fg:bn08}
\end{figure}

\clearpage

\begin{figure}[htbp]
\centering
\includegraphics[width=\textwidth]{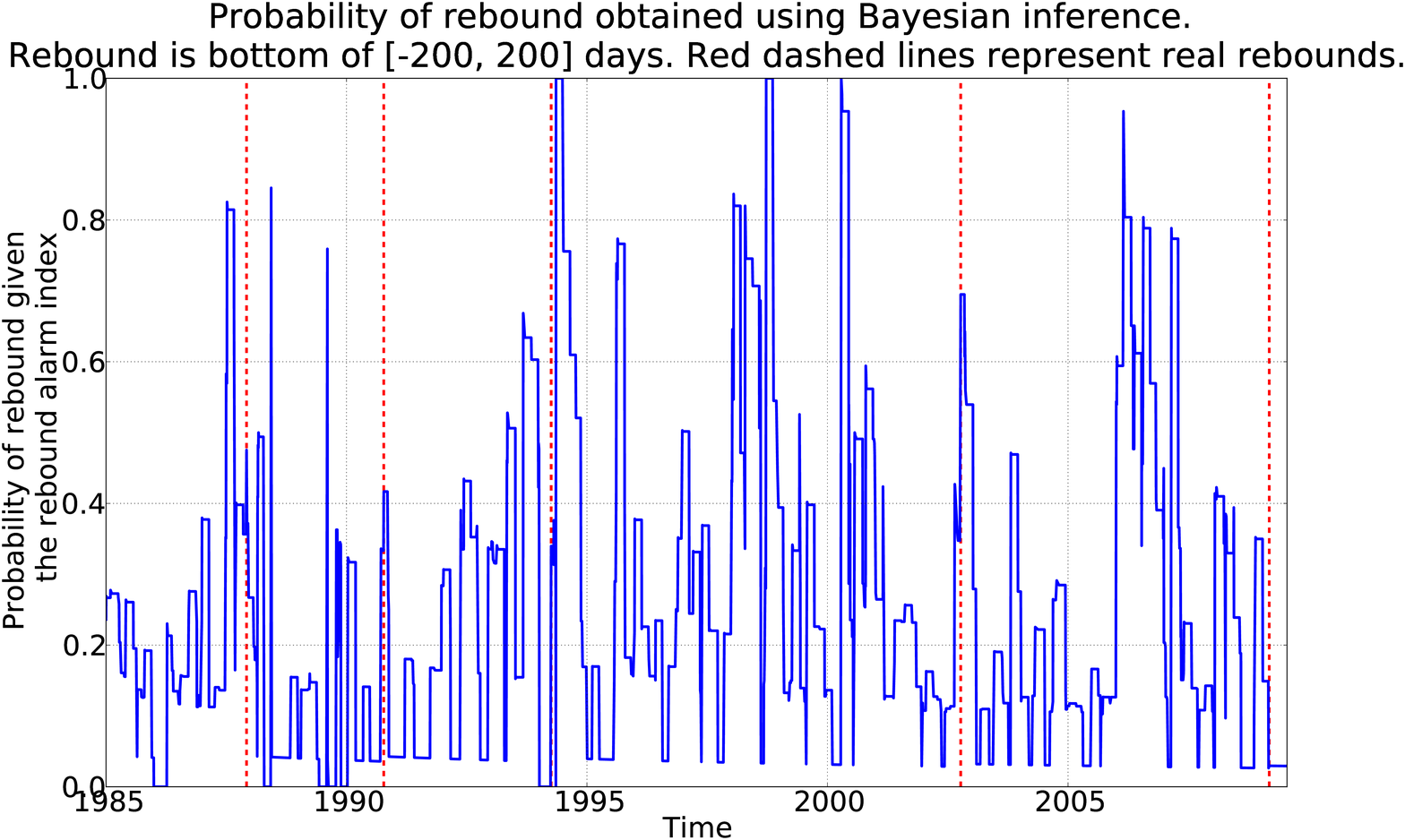}
\caption{Probability of rebound as a function of time $t$, given the
value of the rebound alarm index at $t$, derived by Bayesian inference
 applied to bottoms at the time scale of $\pm 200$ days. The feature qualification is (10, 200). $Lv$
  is the largest rebound index in the past 50 days. The vertical red lines
  show the locations of the realized rebounds in the history of the S\&P500 index.}
\label{fg:bayes}
\end{figure}

\clearpage

\begin{figure}[htbp]
\centering
\includegraphics[width=0.9\textwidth]{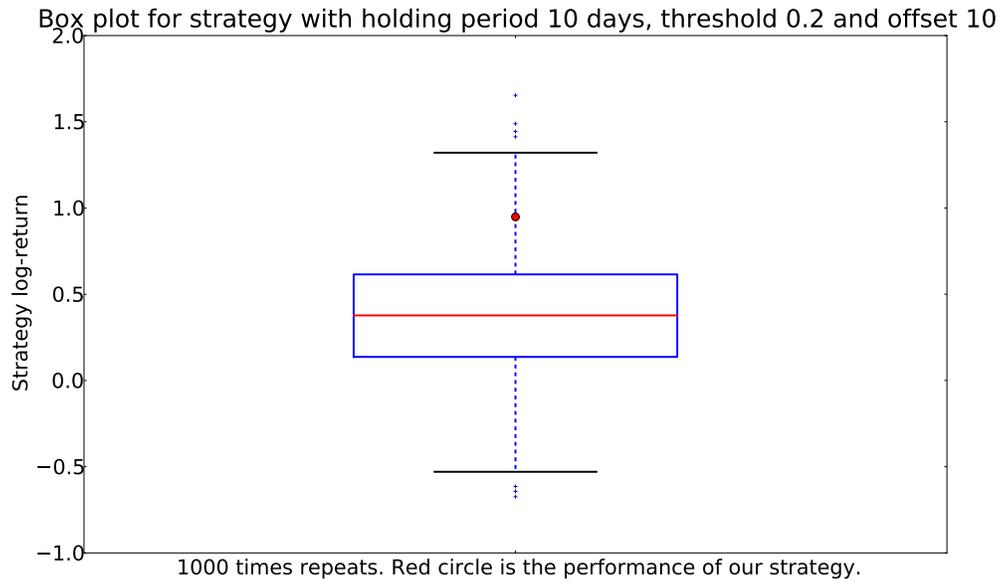}
\caption{Box plot for Strategy I ($Th=0.2,Os=10,Hp=10$). Lower and upper
  horizontal edges (blue lines) of box represent the first and third quartiles.
  The red line in the middle is the median. The lower and upper black lines are
  the 1.5 interquartile range away from quartiles.  Points between quartiles
  and black lines are outliers and points out of black lines are extreme
  outliers.  Our strategy return is marked by the red circle. This shows our
  strategy is an outlier among the set of random strategies.  The log-return ranked 55 out
  of 1000 random strategies.} \label{fg:bp1}
\end{figure}

\begin{figure}[htbp]
\centering
\includegraphics[width=0.9\textwidth]{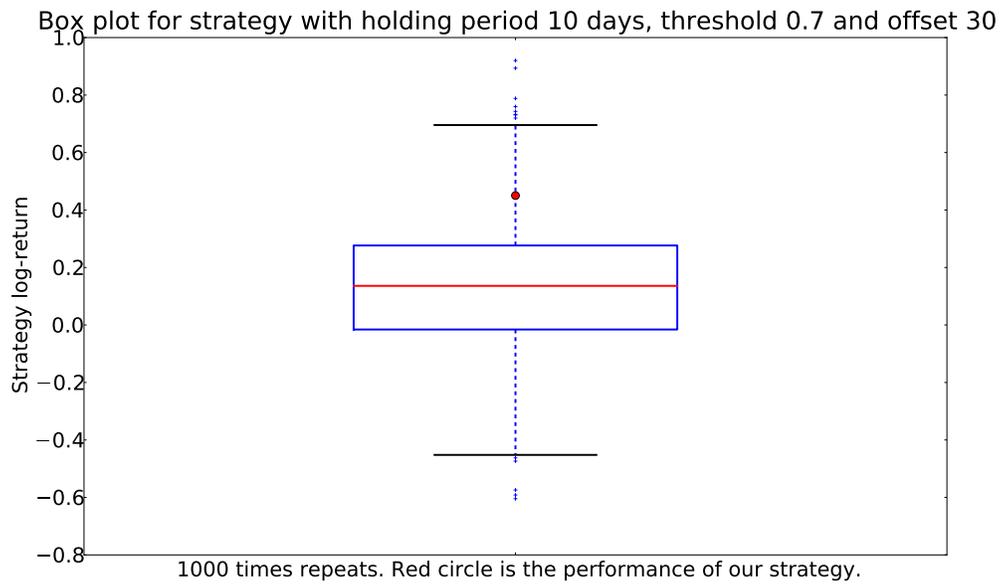}
\caption{Same as figure \protect\ref{fg:bp1} for
Strategy II ($Th=0.7,Os=30,Hp=10$).  The log-return ranked 58 out
  of 1000 random strategies.} \label{fg:bp2}
\end{figure}

\clearpage

\begin{figure}[htbp]
\centering
\includegraphics[width=0.9\textwidth]{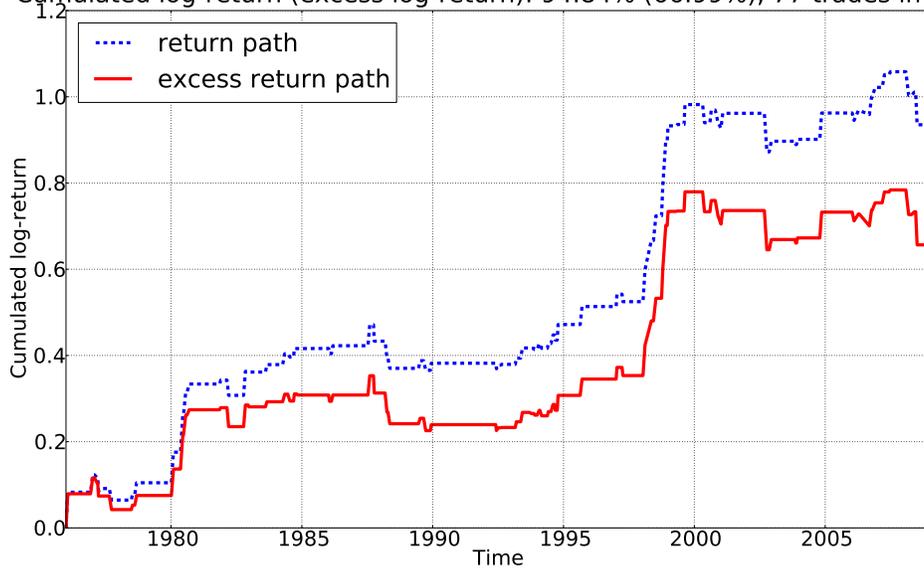}
\caption{ Wealth trajectory for Strategy I ($Th=0.2,Os=10,Hp=10$). Major
  performance parameters of this strategy are: 77 trading times; 66.2\%
  trades have positive return; 1894 total holding days, which is 15.0\% of
  total time. Accumulated log-return is 95\% and average return per trade is
  1.23\%. Average trade length is 24.60 days.  P-value of this strategy is
  0.055} \label{fg:wt1}
\end{figure}

\begin{figure}[htbp]
\centering
\includegraphics[width=0.9\textwidth]{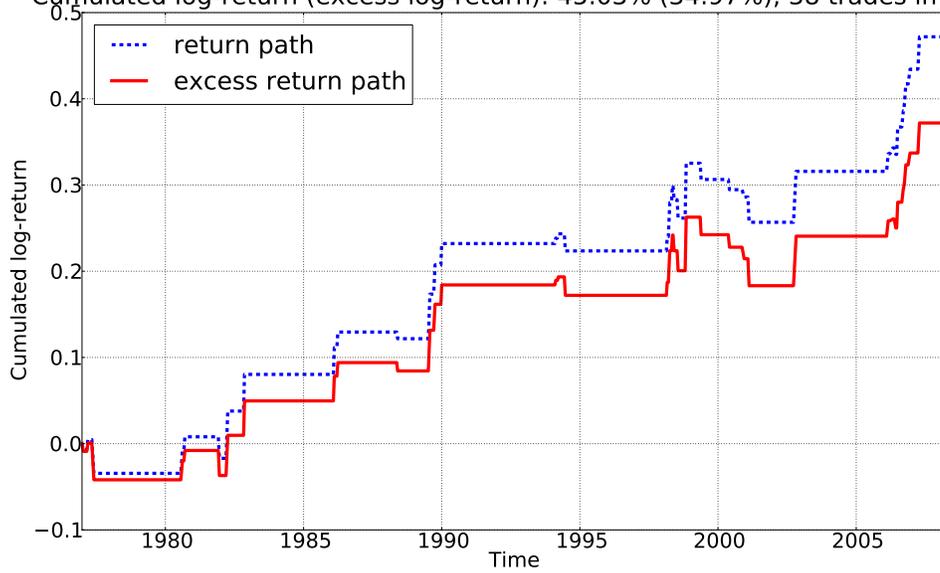}
\caption{ Wealth trajectory for Strategy II ($Th=0.7,Os=30,Hp=10$). Major
  performance parameters of this strategy are: 38 trading times; 65.8\%
  trades have positive return; 656 total holding days, which is 5.2\% of
  total time. Accumulated log-return is 45\% and average return per trade is
  1.19\%. Average trade length is 17.26 days.  P-value of this strategy is
  0.058} \label{fg:wt2}
\end{figure}

\end{document}